\documentclass[aps,twocolumn,prd,showpacs,showkeys,preprintnumbers,superscriptaddress,nobibnotes,floatfix,longbibliography,nofootinbib]{revtex4-1}

\pdfoutput=1

\usepackage[english]{babel}
\usepackage{amsmath,amssymb,amsbsy,amstext,amsthm,amsfonts,array,booktabs,exscale,relsize,slashed,graphicx,upgreek,xcolor,ulem}
\usepackage{multirow,dcolumn,bm,enumerate,url,hyperref}
\usepackage[capitalise]{cleveref}
\usepackage{fontawesome}
\usepackage{mathrsfs}
\usepackage{longtable}
\usepackage{comment}
\definecolor{lightsabergreen}{rgb}{.14,.64,.14}
\definecolor{lightgreen}{rgb}{.14,.44,.14}

\hypersetup{
  colorlinks   = true, %Colours links instead of ugly boxes
  urlcolor     = lightsabergreen, %Colour for external hyperlinks
  linkcolor    = lightsabergreen, %Colour of internal links
  citecolor   = lightsabergreen %Colour of citations
}

\begin{document}
\preprint{
\begin{minipage}{5cm}
\small
\flushright
UCI-HEP-TR-2024-21\\
\end{minipage}} 

\title{Direct Detection of Fast-Moving Low-Mass Dark Matter}

\author{Haider Alhazmi}
\thanks{{\scriptsize Email}: \href{hmalhazmi@jazanu.edu.sa}{hmalhazmi@jazanu.edu.sa}}
\affiliation{Department of Physical Sciences, Jazan University, Jazan 45142, Saudi Arabia}

\author{Doojin Kim}
\thanks{{\scriptsize Email}: \href{doojin.kim@tamu.edu}{doojin.kim@usd.edu}}
\affiliation{Department of Physics, University of South Dakota, Vermillion, SD 57069, USA}
\affiliation{Mitchell Institute for Fundamental Physics and Astronomy, Department of Physics and Astronomy, Texas A\&M University, College Station, TX 77845, USA}

\author{Kyoungchul Kong}
\thanks{{\scriptsize Email}: \href{kckong@ku.edu}{kckong@ku.edu}}
\affiliation{Department of Physics and Astronomy, University of Kansas, Lawrence, KS 66045, USA} 

\author{Gopolang Mohlabeng}
\thanks{{\scriptsize Email}: \href{mailto:gmohlabe@sfu.ca}{gmohlabe@sfu.ca}}
\affiliation{Department of Physics, Simon Fraser University, Burnaby, BC, V5A 1S6, Canada}
\affiliation{TRIUMF, 4004, Westbrook Mall, Vancouver, BC, V6T 2A3, Canada}
\affiliation{Department of Physics and Astronomy, University of California, Irvine, CA 92697, USA}

\author{Jong-Chul Park}
\thanks{{\scriptsize Email}: \href{jcpark@cnu.ac.kr}{jcpark@cnu.ac.kr}}
\affiliation{Department of Physics and Institute of Quantum Systems (IQS), Chungnam National University, Daejeon 34134, Korea}

\author{Seodong Shin}
\thanks{{\scriptsize Email}: \href{sshin@jbnu.ac.kr}{sshin@jbnu.ac.kr}}
\affiliation{Laboratory of Symmetry and Structure of the Universe, Department of Physics, Jeonbuk National University, Jeonju, Jeonbuk 54896, Korea}
\affiliation{Center for Theoretical Physics of the Universe, \\
Institute for Basic Science, Daejeon 34126, Korea}

\date{\today}

\begin{abstract}
We examine the signals produced by dark matter interactions with electrons, which play a crucial role in direct detection experiments employing heavy target materials, particularly in many well-motivated sub-GeV dark matter scenarios. When the momentum transfer to target electrons is comparable to or exceeds their binding energy, atomic effects related to electron ionization become essential for accurately determining signal rates -- especially in the case of fast-moving dark matter.
In this paper, we revisit and extend the atomic ionization formalism, systematically comparing different approaches used to formulate the ionization form factor and identifying their respective domains of validity. As practical applications, we explore detection prospects in xenon target experiments. To illustrate our findings, we consider a specific scenario involving boosted dark matter, which often leads to high-momentum electron recoils. Our analysis demonstrates that the choice of formalism can significantly influence the interpretation of experimental data, depending on the regions of parameter space.
\end{abstract}

\maketitle

%%%%%%%%%%%%%%%%%%%%%%%%%%%%
\section{Introduction}
\label{sec:intro}
%%%%%%%%%%%%%%%%%%%%%%%%%%%%

Dark matter (DM) direct detection experiments have been, for decades, trying to identify the particle nature of DM, mostly through searching for Weakly Interacting Massive Particles (WIMPs). Yet, a clear non-gravitational signal remains elusive~\cite{Jungman:1995df}.
The increasing difficulty of observing WIMPs through nuclear recoils might be explained by their being too light [lighter than $\sim \mathcal O ({\rm GeV})$] to trigger recoils in large-volume detectors.
This problem can be mitigated by (i) searching for electronic recoils where large-volume experiments have altered their readout strategies in order to search for lower mass DM \cite{XENON:2010xwm, Essig:2011nj} or (ii) searching for fast-moving light DM that is moving much faster than the typical Halo DM, thus increasing the energy that it can deposit in a large-volume detector \cite{Agashe:2014yua, Bhattacharya:2014yha, Kong:2014mia, Cherry:2015oca, Necib:2016aez, Alhazmi:2016qcs, Kim:2016zjx, An:2017ojc,
Giudice:2017zke, Chatterjee:2018mej, Kim:2018veo, Cappiello:2018hsu, Bringmann:2018cvk, Ema:2018bih, Kim:2019had, DeRocco:2019jti, Heurtier:2019rkz, Dent:2019krz, Berger:2019ttc, Wang:2019jtk, Kim:2020ipj, Fornal:2020npv, Cao:2020bwd, Jho:2020sku, Alhazmi:2020fju, Guo:2020oum, Dent:2020syp, Jia:2020omh, Jho:2021rmn, Das:2021lcr, Borah:2021yek, Bell:2021xff, An:2021qdl, Bardhan:2022bdg, Maity:2022exk, Bell:2023sdq, Guha:2024mjr, Dutta:2024kuj, Cappiello:2024acu, Choi:2024ism}. 
In this work, we will focus on the latter and fully characterize the electronic recoil signals expected at large-volume direct detection experiments with noble liquid.\\

To gain a better understanding of the anticipated electron recoil signals for fast-moving DM, it is necessary to understand the impact of the associated atomic effects and accurately characterize the atomic ionization functions.
The ionization form factor of a target atom accounts for the probability that an electron with a certain kinetic energy is scattered by DM with a given momentum transfer, depending on the final- and the bound-state wave functions of the electron.
Since direct detection experiments are inclined to use the heavy- or moderate-weight nuclei having complex atomic structures, the precise applications of the form factors play key roles in probing fast-moving DM and, hence, estimating its experimental constraints as well as discovery potential.

The importance of the ionization form factors was emphasized in Refs.~\cite{Essig:2011nj, Essig:2012yx, Roberts:2016xfw, Essig:2017kqs, Catena:2019gfa, Bloch:2020uzh}, in the context of non-relativistic DM.
Some of these studies further highlighted how different ionization form factor formalisms impact the recoil momentum spectrum, particularly at high momentum transfers.
However, much of the literature has largely overlooked this critical aspect, especially in the context of fast-moving DM scenarios proposed to explain the now-subsided Xenon1T electron recoil excess \cite{XENON:2020rca}.
A key point is that many works have utilized the so-called plane wave approximation \cite{Essig:2011nj,Lee:2015qva}, even when analyzing fast-moving DM. Yet, this approach has limitations compared to more advanced methods that compute outgoing electron wavefunctions by solving the Schr\"odinger equation \cite{Essig:2012yx, Essig:2017kqs, Catena:2019gfa, Bloch:2020uzh} or employing the relativistic Dirac-Hartree-Fock method \cite{Roberts:2015lga, Roberts:2016xfw}. These differences become especially pronounced at high momentum transfers, where the effects of fast-moving DM are most relevant. 

In this paper, we investigate the phenomenological implications and impacts of the atomic ionization form factors on the scattering events between fast-moving DM and electrons at direct detection experiments utilizing liquid xenon. 
Fast-moving DM has emerged as a compelling possibility and provides intriguing signals that can be searched for in both DM and neutrino experiments~\cite{Super-Kamiokande:2017dch, COSINE-100:2018ged,  PandaX-II:2021kai, CDEX:2022fig, Super-Kamiokande:2022ncz, CDEX:2022dda, NEWSdm:2023qyb, COSINE-100:2023tcq, PandaX:2024pme}. 
In these scenarios, the main component of DM in the Milky Way is non-relativistic. However, there is some boosting mechanism that induces (semi-)relativistic DM that can reach our detectors on Earth.  
One such mechanism exploits the possibility of a richer dark sector in which DM boost is a generic feature.\footnote{While the goal of this work is to assess the signals induced by general fast-moving DM, for illustration purposes, we will focus on a particular boosting mechanism and leave others for future work.} In this multi-component DM scenario we will refer to as boosted DM (BDM), the heavier DM is the non-relativistic ambient DM in the halo, but can annihilate into lighter (semi-)relativistic states \cite{Belanger:2011ww, Agashe:2014yua}.

For illustration purposes, we focus on xenon-based experiments (similar studies can be carried out for argon-based experiments) and interpret their data by carefully comparing the ionization form factors in the context of BDM scenarios. 
To this end, we employ two benchmark model frameworks: fermionic DM interacting with electrons via an exchange of either i) a vector mediator or ii) a pseudoscalar mediator.\footnote{We note that the theory parameter space for the $s$-wave annihilation scenarios are getting squeezed by cosmological and astrophysical observations when the light DM mass is $\lesssim$ 100 MeV \cite{Kamada:2021muh}. 
On the other hand, the scalar DM whose annihilation is $p$-wave dominant can have more allowed parameter regions. 
In our study, we focus on the impacts of the ionization form factors and will address details of current bounds.}
We show that depending on the model framework, the ionization form factors result in significantly different constraints and sensitivity projections, due to some of the form factors accentuating this model dependence. 

The rest of this paper is organized as follows. In Section~\ref{sec:rates_ion}, we describe the scattering rates between BDM and electrons, including the ionization form factors. 
We briefly review existing treatments of ionization form factors and compare their approaches, including their incorporation of the mediator dependence. 
In Section~\ref{sec:boost}, we briefly discuss a benchmark BDM scenario used for our sensitivity estimates. 
We present and discuss the model-dependent results and interpretations in Section~\ref{sec:candp} and reserve Section~\ref{sec:conc} for discussion and conclusions.

%%%%%%%%%%%%%%%%%%%%%%%%%%
\section{Scattering Rates and Ionization}
\label{sec:rates_ion}
%%%%%%%%%%%%%%%%%%%%%%%%%%

If an incoming DM particle is sufficiently energetic, resulting in a large momentum transfer, it interacts with individual electrons, and the interaction can be described as scattering between DM and free (isolated) electrons. 
However, in the case at hand, the momentum scale is not sufficiently large and the energy is transferred to a target atom, ejecting an electron from it. 
With this detection principle in mind, the event rate of fast-moving DM or BDM (henceforth denoted by $\chi$) scattering with electrons in direct detection experiments is given by

\begin{equation}
    \frac{d R}{d E_r} = N_T \int dE \frac{d\Phi_{\chi}(E)}{dE} \frac{d \sigma_{\chi e}(E) }{d E_r}\,,
    \label{eq:dRdEr}
\end{equation}
%--
where $E_r \equiv E_e^\prime - m_e$ is the electron recoil kinetic energy, $N_T$ is the number of target atoms in the detector, $\sigma_{\chi e}$ is the $\chi$-$e$ scattering cross section, and $E$ is the energy of the incoming BDM. 
Here, $\Phi_\chi$ is the flux of BDM particles entering the detector, and in the most general situations, $\Phi_\chi$ is given by a function of $E$.

Assuming that the incoming DM particle has a mass $m_{\chi}$, energy $E$, momentum $p$, and speed $v$, we find that the differential cross section is given by 
\begin{equation}
\frac{d \sigma_{\chi e} }{d E_r} = \frac{m_{\chi}^2 m_e^2 \overline{\sigma}_{e}}{4 \mu^2 p {k^{\prime}}^2 E E_e v} \int^{q^{+}}_{q^{-}} qdq \, |F_{\rm DM}(q)|^2 \, |f_{\rm ion}(q,E_{r})|^2\,.
\label{eq:dsigma_4}
\end{equation}
The electron energy before scattering (i.e., initial-state energy) is defined in terms of the binding energy BE as $E_e = m_e - {\rm BE}$ which depends on the particular shell in the atom. 
The reduced mass of the DM particle and electron system is defined as $\mu$ and the momentum of the outgoing electron is defined as ${k^{\prime}}$. 
In the differential cross section above, the definite integration is performed over the momentum transfer $q$ ($\equiv |\vec{q}|$, i.e., the magnitude of the three-momentum transfer) with the limits 
\begin{eqnarray}
    q^{\pm} &=& p \pm \sqrt{p^{2}+\Delta E(\Delta E-2 E)} \\
            &=& p \pm \sqrt{p^2 + (E_r+{\rm BE})(E_r+{\rm BE}-2E)}\,,
            \label{eq:qpm}
\end{eqnarray}
where $\Delta E = E_e^\prime - E_e$ is the deposited energy. 
Note that the second row indicates the existence of a range of momentum transfer $q$ corresponding to a given recoil kinetic energy $E_r$. 

The integrand of Eq.~\eqref{eq:dsigma_4} contains two $q$-dependent form factors.
The first form factor is the DM form factor $|F_{\rm DM}(q)|^2$, which characterizes the DM-electron scattering and encapsulates the model dependence. 
This allows us to extract a reference cross section $\overline{\sigma}_{e}$ by redefining the spin-averaged matrix element square $\overline{|\mathcal{M}(q)|}^2$ at a reference value of the momentum transfer $q = \alpha m_{e}$,\footnote{Though we are considering the relativistic DM, choosing this reference value is still a valid assumption since the large momentum transfers in our scenario are encapsulated in the DM form factor.} isolating the $q$-dependent component, following Ref.~\cite{Essig:2011nj}:
\begin{eqnarray}
\overline{|\mathcal{M}(q)|^2} &=& \overline{|\mathcal M (q = \alpha m_e)|^2} \large \times |F_{\rm DM} (q)|^2\,, \nonumber \\
\overline{\sigma}_{e} &=& \frac{\mu^2 \overline{|\mathcal M (\alpha m_e)|^2}}{16\pi m_{\chi}^2 m_e^2}\,.  
\label{eq:DMform}
\end{eqnarray}
The second $q$-dependent form factor of Eq.~\eqref{eq:dsigma_4} is the dimensionless ionization form factor $|f_{\rm ion}(q,E_{r})|^2$ defined in terms of the transition amplitude encapsulating the overlap between the wave functions of the outgoing free electron $\psi_{e_f}$ and that of the initial bound electron $\psi_{e_i}$. 
The treatment of the ionization form factor relies on the estimations of the electron wave functions. 
We consider two treatments, non-relativistic and relativistic treatments.

\subsection{Non-relativistic Treatment}

Here we consider Roothaan-Hartree-Fock (RHF) wave functions for the initial bound state whose radial component $R_{n \ell}$ is given by a linear combination of Slater-type orbitals that are tabulated in Ref.~\cite{Bunge:1993jsz}. 
The two indices $n$ and $\ell$ represent the principal and orbital quantum numbers of the shell at which the initial electron resides. 
For the outgoing electron wave function, we consider two cases.

\medskip

\noindent {\textit{\underline{Case 1}}} is the simplest where the outgoing free electron is considered as a plane wave (PW) \cite{Essig:2011nj, Lee:2015qva, Cao:2020bwd}:
\begin{equation}
|f_{\rm ion}(q,E_{r})|_{{\rm{PW}}}^2 = \frac{(2 \ell + 1) {k'}^{2} }{4 \pi^3{q}} \int d{k} \, {k} |\chi_{n \ell}({k})|^2\,,
\label{eq:PWfion}
\end{equation}
where the integration limit is defined by $|k' \pm q|$. 
The momentum space wave function is related to the radial component of the RHF wave function as
\begin{equation}
\chi_{n \ell}({k}) = 4 \pi i^{\ell} \int dr \, r^2 R_{n \ell}(r) j_{\ell}({k} r)\,, 
\end{equation}
where $j_{\ell}$ is the spherical Bessel function.

\medskip

\noindent {\textit{\underline{Case 2}}} shares the same initial wave function, $R_{n \ell}$, as the previous case, but differs in the estimation of the outgoing wave equation. 
Instead of a simple plane wave for the outgoing free electron, a continuum positive energy solution of the Schrodinger equation of a Coulomb-like potential is considered, $R_{E_r \ell'}$ (see Ref.~\cite{Catena:2019gfa} for detailed $R_{E_r \ell'}$ expressions). 
We abbreviate this case as the Schr\"odinger Continuous Energy (SCE) case for future reference. 
The ionization form factor is given by \cite{Catena:2019gfa}
\begin{eqnarray}
|f_{\rm ion}(q,E_{r})|_{_{\rm{SCE}}}^2 &=& \frac{{k'}^3}{\pi^2} \, \sum_{\ell' = 0}^{\infty} \,\, \sum_{L = |\ell - \ell'|}^{\ell + \ell'} \mathcal{A}^{\ell}_{\ell' L}  \\
&\times& \left| \int_0^{\infty} dr \, r^2 j_{L}({q}r) R^{*}_{E_r \ell'}(r) R_{n \ell}(r) \right|^{2}\,. \nonumber
\label{eq:SCE}
\end{eqnarray}
Here, the overall coefficient $\mathcal{A}^{\ell}_{\ell' L}$ is given in terms of the Wigner 3j-symbol by
\begin{equation}
\mathcal{A}^{\ell}_{\ell' L} \equiv \left(2 \ell+1\right) \left(2 \ell'+1\right) (2 L+1) \left(\begin{array}{ccc}
\ell & \ell' & L \\
0 & 0 & 0
\end{array}\right)^2\,.
\label{eq:Acoeff}
\end{equation}

We note that similar formalisms for the ionization form factor based on the Schrodinger treatment have been developed in Refs.~\cite{Essig:2012yx, Essig:2015cda, Essig:2017kqs, Hamaide:2021hlp}.
A comparison between these formalisms was conducted in Ref.~\cite{Emken:2024nox}. 
The differences in the interpretation of experimental data across these formalisms are on the order of tens of percent. 
Therefore, in the rest of this paper, we use the formalism from Ref.~\cite{Catena:2019gfa}.

\subsection{Relativistic Treatment} 
\label{sec:reltreat}

The relativistic treatment has the flexibility to accommodate various Lorentz interaction structures. 
In this scenario, both the bound and free wave functions are taken as solutions to the Dirac equation instead of the Schr\"{o}dinger equation. 
The bound state corresponds to the negative quantized energy solutions, while the free state corresponds to the positive continuous energy solution; therefore, we choose to abbreviate this case by the Dirac Continuous Energy (DCE) case. 
Assuming that electrons are bound by a spherically symmetric effective potential $V_{\rm eff}(r)$, we have the relativistic Dirac Hamiltonian, 
\begin{equation}
    \hat{H}=\vec{\alpha}\cdot \hat{\vec{p}}+m_e(\beta-1)+V_{\rm eff}(r)\,,
\end{equation}
where $\vec{\alpha}$ and $\beta$ are the usual Dirac matrices in the standard representation. 
Since the potential is only radially dependent, the general bound-state solution $\psi_{n\kappa m}$ to the eigenvalue problem $\hat{H}\psi_{n\kappa m}=E_{n\kappa}\psi_{n\kappa m}$ depends on the radius only: 
\begin{equation}
    \psi_{n\kappa m}(\vec{r})=\frac{1}{r}
    \begin{pmatrix}
    P_{n\kappa}(r)\Omega_{\kappa m}(\hat{r}) \\
    iQ_{n\kappa}(r)\Omega_{-\kappa m}(\hat{r})
    \end{pmatrix}\,,
\end{equation}
where $\Omega$ denotes the two-component spherical spinor and $P_{n\kappa}$ and $Q_{n\kappa}$ denote the ``large'' and ``small'' Dirac radial components. 
For the continuum-state solutions, $P_{n\kappa}$ and $Q_{n\kappa}$ are replaced by $P_{E_r\kappa}$ and $Q_{E_r\kappa}$, respectively.

One can show that the ionization form factor is independent of the angular component of the wave functions (i.e., $\Omega_{\kappa m}$) and only depends on the radial parts $P$ and $Q$. 
Note that the bound-state radial parts $P_{n \kappa}$ and $Q_{n \kappa}$ are characterized by two quantum numbers, the principal quantum number $n$ and the Dirac quantum number $\kappa$, while the free state radial parts $P_{E_r \kappa'}$ and $Q_{E_r \kappa'}$ are characterized by the energy $E_r$ in addition to the Dirac quantum number $\kappa'$. 
The Dirac quantum number is defined in terms of the orbital ($\ell$) and total ($j$) angular momentum quantum numbers as 
\begin{equation}
    \kappa =  (\ell - j)(2j +1)\,.
\end{equation}
The ionization form factor is given in terms of the $q$-dependent transition factor, $|f_{e_{i} \rightarrow e_{f}}|^2$, as
\begin{equation}
|f_{\rm ion}(q,E_{r})|^2 = \frac{4 {k^{\prime}}^3}{(2 \pi)^{2}} \, |f_{e_{i} \rightarrow e_{f}}|^2\,,
\label{eq:trans}
\end{equation}
where $e_i$ and $e_f$ respectively correspond to the initial and final electron states. 
In this work, we will consider two types of interactions, vector and pseudoscalar interactions. 
For completeness, however, we provide the transition factor expressions for vector (V), axial vector (A), scalar (S), and pseudoscalar (P) interactions as follows \cite{Roberts:2016xfw}:
\vspace*{-0.1cm}
\begin{equation}
|f_{e_{i} \rightarrow e_{f}}|_{_{\rm{DCE, V}}}^2 = \sum_{\kappa'} \,\sum_{L} \, C_{\kappa \kappa'}^{L} \, (R_{P,P} + R_{Q,Q})^2\,, \label{eq:relFFdiffmed} 
\end{equation}
\vspace*{-0.3cm}
\begin{equation}
|f_{e_{i} \rightarrow e_{f}}|_{_{\rm{DCE, A}}}^2 = \sum_{\kappa'} \,\sum_{L} \, D_{\kappa \kappa'}^{L} \, (R_{P,Q} - R_{Q,P})^2\,, 
\end{equation}
\vspace*{-0.3cm}
\begin{equation}
|f_{e_{i} \rightarrow e_{f}}|_{_{\rm{DCE, S}}}^2 = \sum_{\kappa'} \,\sum_{L} \, C_{\kappa \kappa'}^{L} \, (R_{P,P} - R_{Q,Q})^2\,, 
\end{equation}
\vspace*{-0.3cm}
\begin{equation}
|f_{e_{i} \rightarrow e_{f}}|_{_{\rm{DCE, P}}}^2 = \sum_{\kappa'} \,\sum_{L} \, D_{\kappa \kappa'}^{L} \, (R_{P,Q} + R_{Q,P})^2\,, \label{eq:relFFdiffmedP}
\end{equation}
where the $R$ factors in the brackets above are the radial integrals defined as 
\begin{equation}
R_{X, Y} \equiv \int_0^{\infty} \, X_{E_r \kappa'} Y_{n \kappa} j_L(q r) dr\,. \label{eq:radialdef}
\end{equation}
Here the indices $X$ and $Y$ are replaced with their corresponding ones in Eqs.~\eqref{eq:relFFdiffmed} to~\eqref{eq:relFFdiffmedP}. 
The overall coefficient $C_{\kappa \kappa^{\prime}}^L$ is written in terms of the Wigner $3j$-symbol as     
\begin{widetext}
\begin{equation}
    \begin{aligned}  C_{\kappa \kappa^{\prime}}^L&=\frac{1}{4}(-1)^{j+j^{\prime}-{\ell}-l^{\prime}}(2 L+1)\left(\begin{array}{ccc}{\ell}^{\prime} & {\ell} & L \\ 0 & 0 & 0\end{array}\right)^2\left(\begin{array}{ccc}j^{\prime} & L & j \\ -\frac{1}{2} & 0 & \frac{1}{2}\end{array}\right)^{-2}\left[(-1)^{j+j^{\prime}-{\ell}-{\ell}^{\prime}}(2 j+1)\left(2 j^{\prime}+1\right)\left(\begin{array}{ccc}{\ell}^{\prime} & {\ell} & L \\ 0 & 0 & 0\end{array}\right)^2\right. \\ &\left.+\,8 \sqrt{{\ell}^{\prime}\left({\ell}^{\prime}+1\right) {\ell}({\ell}+1)}\left(\begin{array}{ccc}{\ell}^{\prime} & {\ell} & L \\ 0 & 0 & 0\end{array}\right)\left(\begin{array}{ccc}{\ell}^{\prime} & {\ell} & L \\ -1 & 1 & 0\end{array}\right)-4\left(\kappa^{\prime}+1\right)(\kappa+1)\left(\begin{array}{ccc}{\ell}^{\prime} & {\ell} & L \\ -1 & 1 & 0\end{array}\right)^2\right]\,.\end{aligned}
\end{equation}
\end{widetext}
The coefficient $D_{\kappa \kappa^{\prime}}^L$ has a similar form, but can also be found through simple replacements in the indices of the above coefficient; it can be obtained using $C_{\kappa\kappa'}^L$ with $\kappa$ and $\ell$ replaced by $-\kappa$ and $|1/2-\kappa|-1/2$, respectively. 
For a detailed discussion on the numerical stability of the above integrals and the methods used to estimate the relativistic wave functions, see e.g., Refs.~\cite{Roberts:2015lga,Roberts:2016xfw, Alhazmi:2021qgd}.

As we observe from Eqs.~\eqref{eq:relFFdiffmed} through \eqref{eq:relFFdiffmedP}, the ionization form factors vary depending on the mediator type. 
For example, the vector interactions do not mix large and small parts of the Dirac wave function, whereas the pseudoscalar interactions do. 
Hence, it is more straightforward for the ionization form factors derived from the relativistic treatment to display model dependence. 
Achieving model dependence in the SCE-based treatment is also possible, but less straightforward.
Reference \cite{Catena:2019gfa} demonstrated that the entire framework, including the matrix element squared which is treated non-relativistically, can have some non-trivial model dependence quantified through an effective field theory approach. 
We found it non-trivial to map the model structure of the form factors from the approach in Ref.~\cite{Catena:2019gfa} to our relativistic treatment, hence we only incorporate what they refer to as the scalar form factor (we call this the SCE treatment). 
We also note that this is readily incorporated into models of fermionic DM with a vector mediator, as has been considered in great detail in the literature.

\subsection{Comparisons and Discussions}

\begin{figure*}[t!]
\centering
\hspace{-.5cm} \includegraphics[width=.34\textwidth]{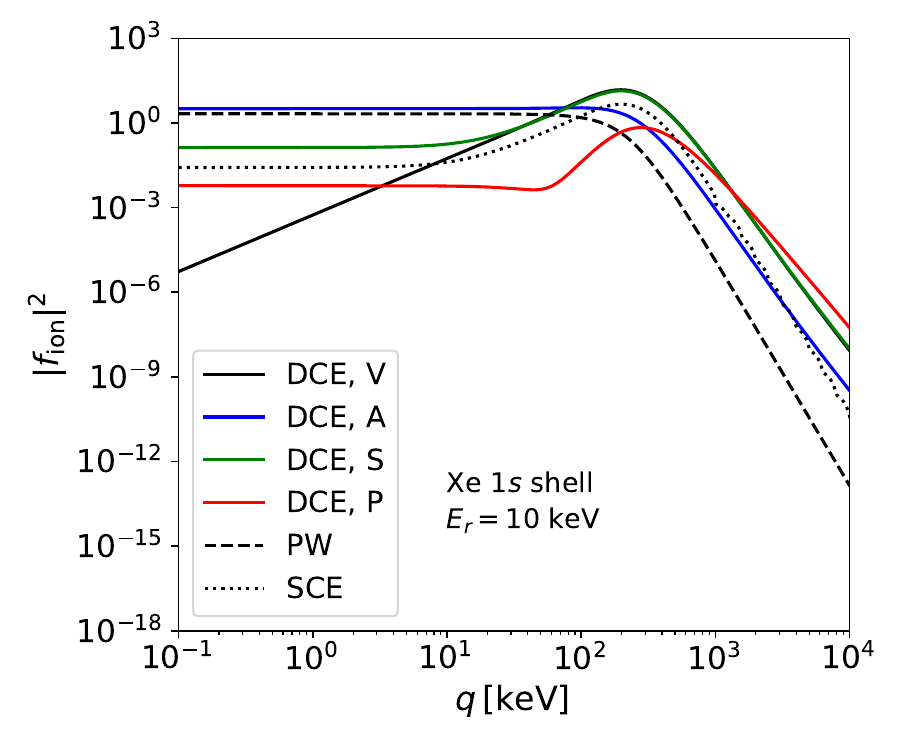}
\hspace{-.2cm} \includegraphics[width=.34\textwidth]{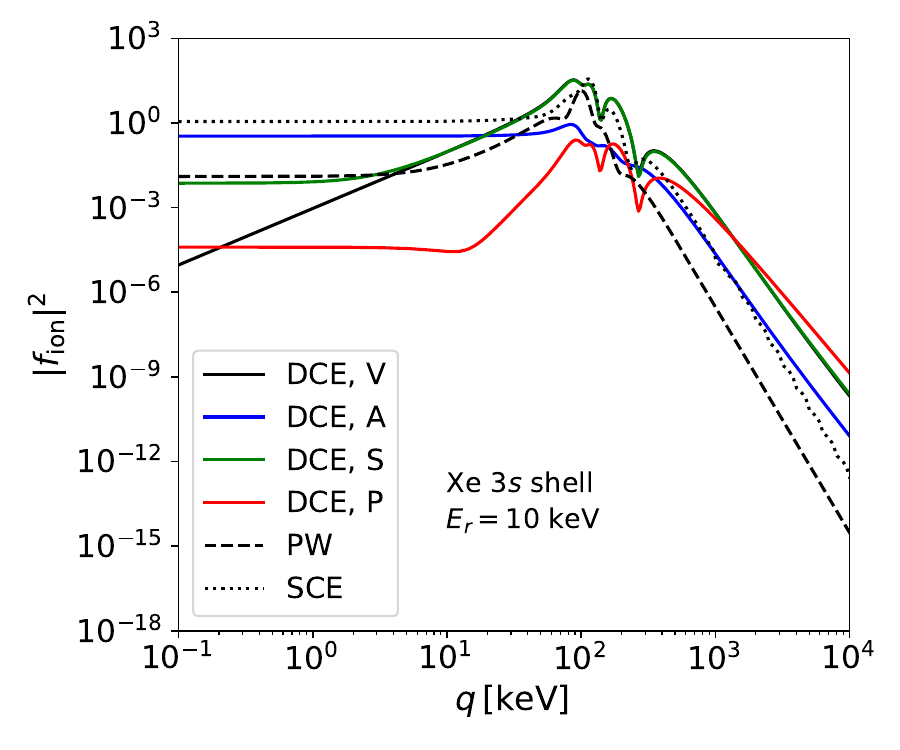} 
\hspace{-.2cm} \includegraphics[width=.34\textwidth]{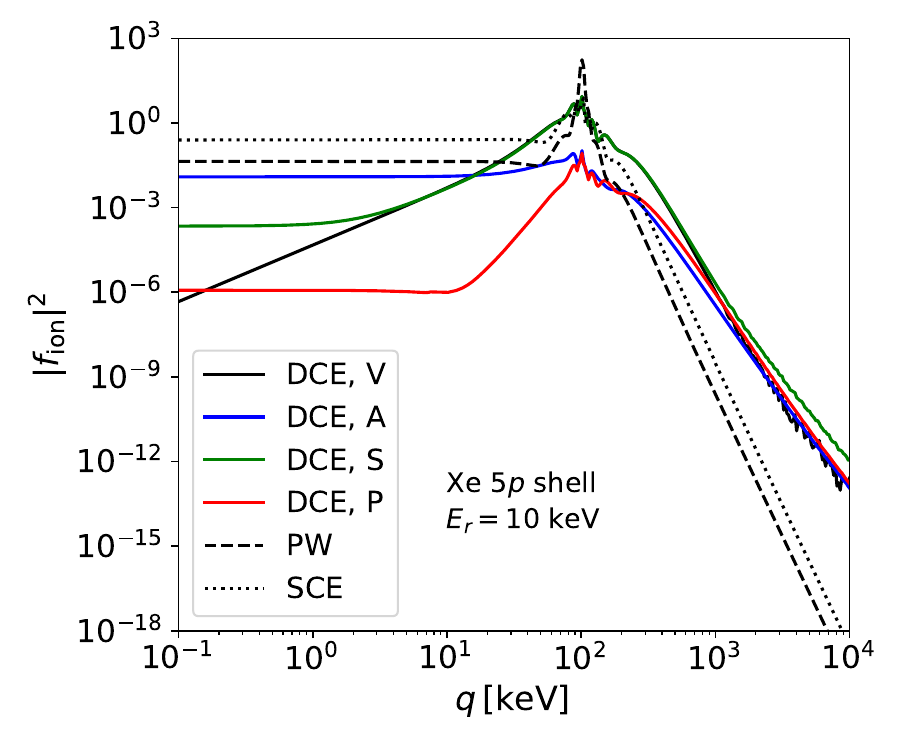}
\caption{Comparisons of ionization form factors for Xe 1s shell (left), i.e., $n=1$ and $\ell=0$, for Xe 3s shell (center), i.e., $n=3$ and $\ell=0$, and for Xe 5p shell (right), i.e., $n=5$ and $\ell=1$, at fixed recoil kinetic energy of 10 keV. 
The dotted and dashed lines are the results for the cases with the non-relativistic PW and SCE treatments, respectively. 
The solid lines are the results of the relativistic treatment where black, blue, green, and red represent the vector, axial vector, scalar, and pseudoscalar interactions, respectively. 
\label{fig:Ar_vs_Xe_ion} }
\end{figure*}

We are now ready to compare our ionization form factors derived using the relativistic approximation with those obtained from non-relativistic treatments. Figure~\ref{fig:Ar_vs_Xe_ion} displays three illustrative comparisons of various ionization form factors for xenon. In the left panel, we show the 1s shell, the 3s shell in the middle panel, and the outermost shell (5p) in the right panel. 
All plots are illustrated at a fixed recoil kinetic energy of 10 keV, comparing the different ionization form factor formalisms discussed above. 
The solid black, green, blue, and red lines represent the relativistic vector, scalar, axial vector, and pseudoscalar form factors, respectively. 
The dashed and dotted lines correspond to the PW and SCE formalisms, respectively.
In all three panels, we notice an exponential suppression in the non-relativistic treatment at high $q$. 
This is due to the suppression of the electron wave function at short distances from the nucleus. 
In contrast, in the relativistic framework, particularly one utilizing the relativistic Hartree-Fock method, this suppression is alleviated because ionization is predominantly governed by the contribution from s-wave electrons at small electron-nucleus distances. 
For a more detailed explanation, see Refs.~\cite{Roberts:2016xfw,Roberts:2015lga}.
Note that we show the ionization form factor in the wide range of the momentum transfer for illustration and shape comparison, while the minimum and maximum momentum transfers $q^\pm$ associated with the given $E_r$ and electron shell are determined for a given DM four-momentum as in Eq.~\eqref{eq:qpm}.

It is important to understand the effects of the ionization form factor when combined with the matrix element squared. 
As noted in Eq.~\eqref{eq:dsigma_4}, the differential cross section contains two $q$-dependent factors: the DM form factor and the ionization form factor. 
While the general behavior of the ionization form factor follows that shown in Figure~\ref{fig:Ar_vs_Xe_ion}, its impact on the final result is shaped by its interplay with the DM form factor, resulting in varying outcomes, depending on the specific model. 
To fully understand the combined effect, it is necessary to analyze the $q$-dependence within the complete differential scattering cross section.\footnote{Although we will discuss the details about our benchmark BDM scenarios in Section~\ref{sec:boost}.} 

\begin{figure*}[t!]
\begin{center}
\hspace{0.0cm} \includegraphics[width=.45\textwidth]{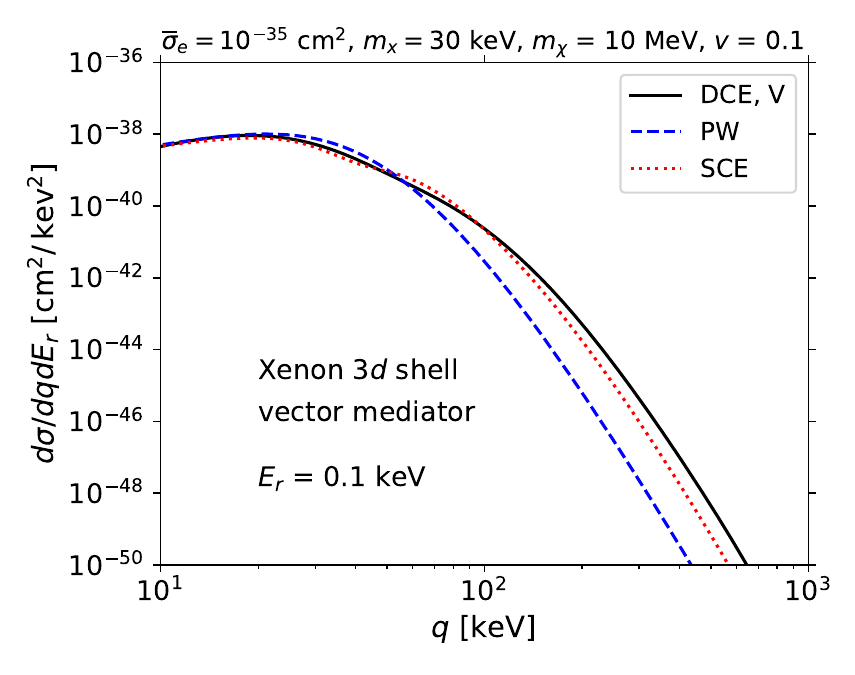}
\hspace{0.0cm} \includegraphics[width=.45\textwidth]{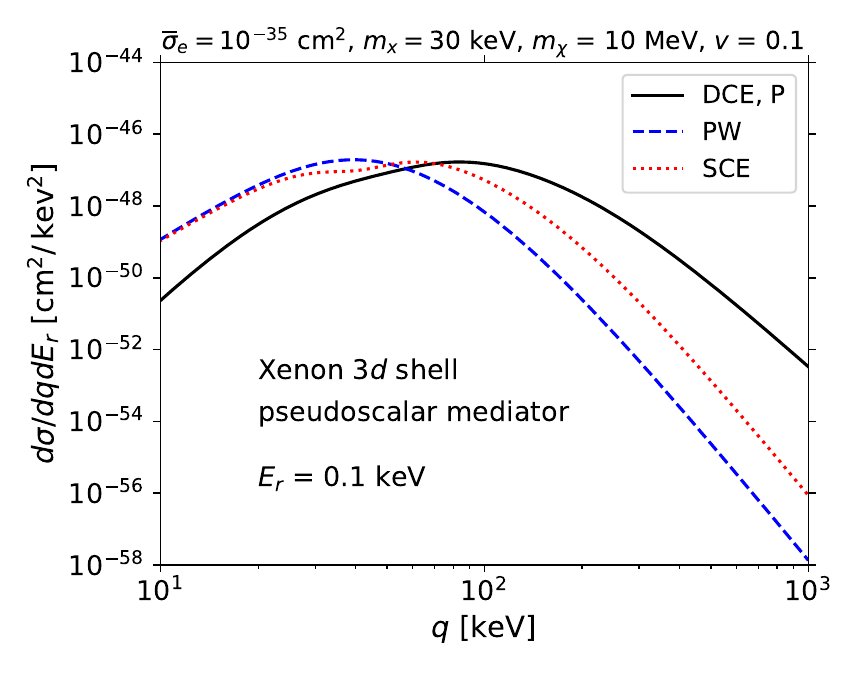}\\
\hspace{0.0cm} \includegraphics[width=.45\textwidth]{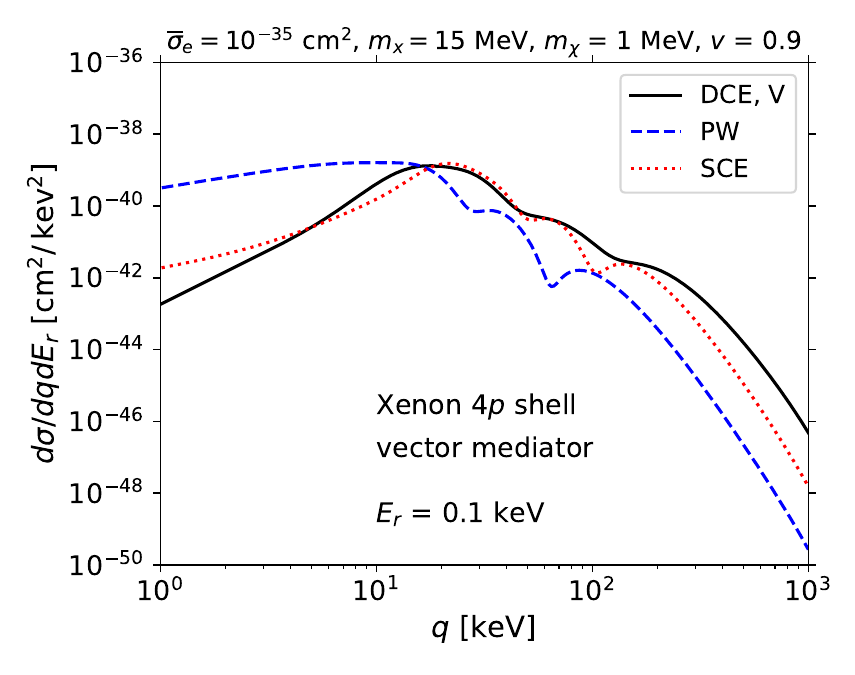}
\hspace{0.0cm} \includegraphics[width=.45\textwidth]{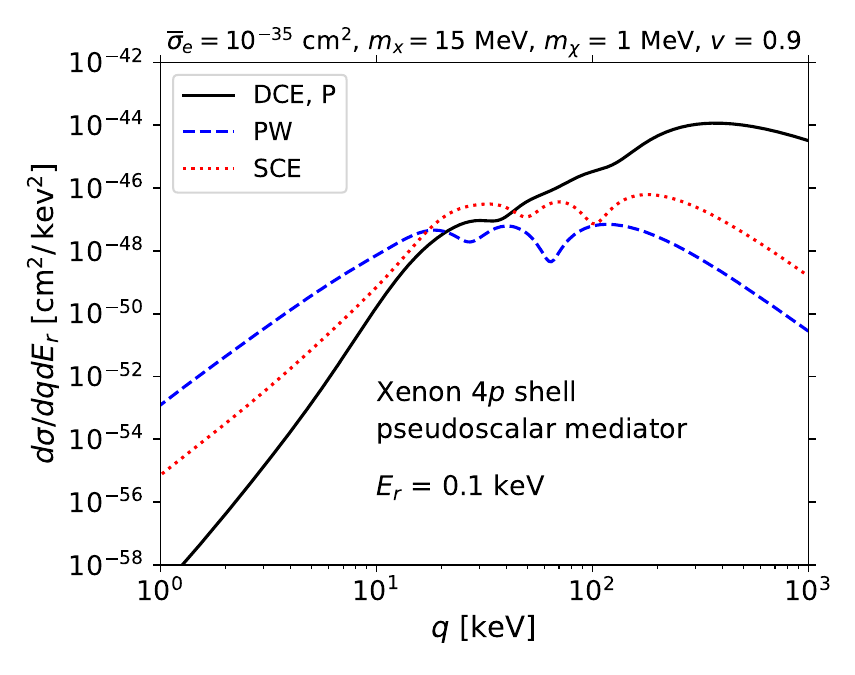}\\
\end{center}
\vspace*{-0.5cm}
\caption{Differential cross section comparison between the relativistic (black solid lines) and the non-relativistic (blue dashed and red dotted lines) treatments for the vector mediator (left panels) and the pseudoscalar mediator (right panels). 
With $X$ collectively denoting the mediators, the top panels correspond to the lighter mediator with $m_X= 30$ keV and  $v=0.1c$ at the xenon $3d$ shell. 
The bottom panels correspond to the heavier mediator with $m_X= 15$ MeV and $v=0.9c$ at the xenon $4p$ shell.}
\label{fig:FionFDM}
\end{figure*}

In Fig.~\ref{fig:FionFDM}, we quantify the effect of the ionization form factors on the differential scattering cross section for the different model frameworks that we consider. 
In the upper-left panel, we plot the vector mediator case for the 3d shell. 
We see that the relativistic (DCE) and non-relativistic (PW and SCE) treatments result in similar-shaped distributions. 
Both treatments peak at the same location and exhibit a similar declining trend, with the relativistic treatment dominating at high momentum transfer. 
In contrast, as shown in the right panel, for the pseudoscalar mediator, there are distinct differences between the two treatments. 
The relativistic (DCE) treatment peaks at higher momentum transfer, due to the $q$ dependence of the matrix element, and displays a rising behavior in regions where the non-relativistic (PW and SCE) treatment shows a declining trend. 
Eventually, all curves asymptotically fall at very high momentum transfer. 
These observations are demonstrated at different benchmark values for the DM particle mass and the mediator mass, and different energy shells of the xenon atom (as seen in the lower panels, which show similar trends).
We choose the 3d and 4p shells for illustrative purposes, as we can see similar trends in other shells.
The chosen $q$-scale in this figure adheres to the range specified in Eq.~\eqref{eq:qpm}. 
The minimum expected momentum transfer is $q \gtrsim$ 10 keV for the choice of parameters in Fig.~\ref{fig:FionFDM}.

%%%%%%%%%%%%%%%%%%%%%%%%%%
\section{Benchmark BDM Scenarios}
\label{sec:boost}
%%%%%%%%%%%%%%%%%%%%%%%%%%

For illustration, we envision the model of two-component DM~\cite{Belanger:2011ww} as a boosting mechanism where the heavier DM species, say $\chi_0$, which dominates over the halo density, pair-annihilates to the lighter DM species $\chi$ in the present universe, i.e., $\chi_0\bar\chi_0 \to \chi\bar\chi$. 
Therefore, the energy of the incoming BDM $\chi$ is simply monochromatic, the same as the mass of $\chi_0$, i.e., $E=m_{\chi_0}$. 
The flux of incoming DM particles across the galaxy is given by \cite{Agashe:2014yua}
\begin{eqnarray}
    \frac{d\Phi_\chi}{dE}&=&1.6 \times 10^{-4}{\rm cm}^{-2}{\rm s}^{-1}  \\ 
    &\times& \left (\frac{\langle \sigma v \rangle_{\chi_0\bar\chi_0\rightarrow \chi\bar\chi}}{5 \times 10^{-26}{\rm cm}^3{\rm s}^{-1}}\right ) \left ( \frac{\rm GeV}{m_{\chi_0}}\right )^2\delta (E-m_{\chi_0})\,, \nonumber
\end{eqnarray}
where $\langle \sigma v \rangle_{\chi_0\bar\chi_0\rightarrow \chi\bar\chi}$ is the velocity-averaged annihilation cross section. 
With this choice, the differential event rate in Eq.~\eqref{eq:dRdEr} can be greatly simplified. 
We do not elaborate further on the details of the annihilation process, as the flux depends solely on the annihilation cross section and the mass of $\chi_0$. 

To describe the detection of BDM-induced signals, it is necessary to specify the interactions between the BDM and electrons. 
As concrete examples, we here focus on two benchmark scenarios in which the incoming fermionic BDM produces electron recoils through a $t$-channel exchange of either a vector or a pseudoscalar mediator. 
The relevant interaction Lagrangians are given by
\begin{eqnarray}
    -\mathcal{L}_{\rm V, int} &\supset& g_D V_\mu\bar{\chi}\gamma^\mu\chi +g_e  V_\mu\bar{e}\gamma^{\mu}e\,, \\
    -\mathcal{L}_{\rm PS, int} &\supset& i c_D a\bar{\chi}\gamma^5\chi +i c_e a\bar{e}\gamma^5 e\,, 
\end{eqnarray}
where $g_D$ and $c_D$ represent the couplings within the dark sector while $g_e$ and $c_e$ parameterize the coupling strengths between the electron and the mediators. 
Under the interactions defined here, the spin-averaged matrix elements as a function of $q$ are given by 
\begin{eqnarray}
   \overline{|\mathcal{M}|}_{\rm V}^2&=&  \frac{2g_D^2g_e^2}{(t-m_V^2)^2} \left\{ 2( m_\chi^2+ m_e^2)^2 \right. \\
    &&- \left. 4( m_\chi^2+m_e^2)s +2 s^2+2 s t+t^2 \right\}\,, \nonumber \\
    \overline{|\mathcal{M}|}_{\rm PS}^2&=&\frac{c_D^2c_e^2}{(t-m_a^2)^2}t^2\,,
\end{eqnarray}
where Mandelstam variables $s$ and $t$ are given by 
\begin{eqnarray}
s&=& m_\chi^2+2E(m_e-{\rm BE})+ (m_e-{\rm BE})^2\,, \\
t&=&(\Delta E)^2-q^2\,.
\end{eqnarray}

%%%%%%%%%%%%%%%%%%%%%%%%%%
\section{Constraints and Prospects}
\label{sec:candp}
%%%%%%%%%%%%%%%%%%%%%%%%%%

\begin{table*}[t!]
    \centering
    \renewcommand{\arraystretch}{1.4} % Increases row height
    \begin{tabular}{|c|c|c|c|}
        \hline
        Experiment & XENONnT~\cite{XENON:2024wpa, XENON:2022ltv} & ~~~XENON1T~\cite{XENON:2017lvq,XENON:2018voc}~~~ & ~~~XENON100~\cite{XENON100:2011cza, XENON:2016jmt}~~~ \\ \hline
        Electron Recoil & ~$1$ keV $ < E_r < 140 $ keV~\cite{XENON:2022ltv} &  ~$14 < {\rm PE} < 140 $~\cite{XENON:2021qze}  & ~$ 80 < {\rm PE} < 10^3 $~\cite{XENON:2016jmt} \\ \hline
        ~Exposure [ton$\cdot$year]~ & $1.16$~~\cite{XENON:2022ltv}  & $6 \times 10^{-2}$~\cite{XENON:2021qze} & $3 \times 10^{-2}$~\cite{XENON:2016jmt}  \\ \hline
        Analysis & $E_r$ spectrum & PE spectrum  & PE spectrum  \\ \hline
    \end{tabular}
    \vspace*{0.2cm}
    \caption{Summary of experiments, including the considered recoil energy range for XENONnT, as well as the photoelectron range for XENON1T and XENON100. Additionally, the exposure is provided in ton$\cdot$years, and the final row specifies the type of signal-background analysis conducted.}
    \label{tab:EXPs}
\end{table*}

We are now in the position to study the constraints and sensitivity reaches of our benchmark BDM scenarios, highlighting the importance of the ionization form factor effects.
To this end, we consider Xe-based DM direct detection experiments, XENON100~\cite{XENON100:2011cza, XENON:2016jmt}, XENON1T~\cite{XENON:2017lvq,XENON:2018voc}, and XENONnT~\cite{XENON:2024wpa, XENON:2022ltv},  although the analysis scheme can be directly applied to other noble liquid experiments.

\medskip

\noindent i) {\bf XENONnT} is a multi-ton xenon-based experiment primarily designed to study WIMP interactions with target nuclei, while also providing effective capabilities for investigating electron recoils. 
In our analysis, we use the differential rate from Eq.~\eqref{eq:dRdEr} to calculate the signal rate, incorporating efficiency corrections, selections, and binning to enable a direct comparison with the measured rate reported by the XENONnT collaboration~\cite{XENON:2022ltv}, based on an exposure of 1.16 ton-years. 
The study of BDM at XENONnT and similar experiments focuses on the energy spectrum in the tens of keV range, enabling the detection of moderately boosted particles and probing BDM at relatively low energies.

\medskip

\noindent ii) {\bf XENON1T} has similar capabilities for studying BDM, with an exposure of 0.06~ton$\cdot$years. 
Although the exposure is relatively small, XENON1T serves as an example for investigating electron recoils at much lower energies, on the order of a keV or below. To study BDM in this energy range, we first convert the signal electron recoil spectrum into an electron yield, $n_e$, and then into a photoelectron (PE) distribution. The electron yield arises because the initially scattered electron in the target is energetic enough to cause ionization and excitation in surrounding atoms, producing observable quanta. For the electron yield, we follow the modeling outlined in Ref.~\cite{Essig:2012yx}. Subsequently, $n_e$  generates a PE distribution, representing the observed scintillation light in the detector. We adopt a Gaussian modeling approach similar to that used by the XENON1T experiment, where the mean is approximated as $n_e G$ and the width as $\sqrt{n_e}\Delta G$. Here, $G$ and $\Delta G$ denote the scintillation gain and its spread, respectively, with numerical values of $G=28.8$ PE and $\Delta G=7.13$ PE. Once the PE distribution is determined, we bin the data and apply detector acceptance criteria to enable a direct comparison with the measured events reported by XENON1T \cite{XENON:2021qze}.

\medskip 

\noindent iii) {\bf XENON100} has an exposure of $0.03$ ton$\cdot$years, and we apply a similar treatment as described for the XENON1T experiment. The PE distribution is also modeled with a Gaussian, but in this case, we use the numerical values $G=19.7$ PE and $\Delta G=6.2$ PE, following Ref.~\cite{Essig:2017kqs}. We apply binning, detector acceptance, and detector efficiency to the PE distribution and then compare the resulting signal with the measured events reported in Ref.~\cite{XENON:2016jmt}.

\medskip

Table \ref{tab:EXPs} summarizes these experiments along with their corresponding energy spectrum or photoelectron range, exposure, and type of analysis. For instance, the signal sensitivity analysis XENONnT is performed over the electron recoil spectrum compared to the binned spectrum reported by the experiment. Similarly, XENON1T and XENON100 follow the approach of XENONnT, but the analysis is focused on the photoelectron spectrum produced by recoil electrons.

\medskip

For these experiments, we evaluate signal sensitivity using the exclusion significance, defined as
\begin{equation}
    \sigma_{\rm exc} = \sqrt{2 \sum_{i=1} ^{N} \left[ B_i \ln \left( \dfrac{B_i}{B_i + S_i}\right) + S_i \right]},
    \label{eq:significance}
\end{equation}
where $S$ and $B$ represent the BDM signal and the measured background for the experiment of interest, respectively. The index $i$ sums over the bins in cases where the background is binned. Our results are presented as 2$\sigma$ or 95\% confidence limit plots on the BDM parameter space. This study does not aim to establish the best experimental limits on the BDM parameter space but rather to highlight the dependence of signal sensitivity estimates on ionization form factor effects. 

Our free parameters are the reference cross section $\overline{\sigma}_e$, the mediator mass $m_X$ (i.e., either $m_V$ or $m_a$ depending on the mediator type), the speed $v$ and the mass $m_{\chi}$ of the BDM particle.
Exclusion limit plots are presented in the two-dimensional parameter space of ($m_{\chi}$, $\overline{\sigma}_e$).
As a benchmark, we compare a relatively heavy mediator mass of $m_X = 15$ MeV to a lighter mediator mass of $m_X = 30$ keV, while varying the speed by selecting high boost ($v = 0.999$), moderate boost ($v = 0.9$), and low boost ($v = 0.1$).
First, we will begin by exploring the effects of different treatments outlined in Sec.~\ref{sec:rates_ion}, demonstrating how these effects can vary significantly depending on the mediator type: vector and pseudoscalar. For illustration, we will highlight these effects at high DM speed and then compare them to a much lower speed at both XENON1T and XENONnT. Finally, we will present summary limit plots considering relativistic treatment only for all the aforementioned experiments and benchmark speed values.

We note that in our analysis we did not include the Earth attenuation effects as considered in Ref.~\cite{Cappiello:2018hsu}, although a simplified estimation was made in our earlier work~\cite{Alhazmi:2020fju}.
Our focus here is electron scattering and any attenuation by the overburden is non-trivial (see Ref.~\cite{Emken:2019tni} for attenuation of non-relativistic DM) and is beyond the scope of this work. 
Nevertheless, when presenting the sensitivity estimates later on, we will indicate the reference cross section $\bar{\sigma}_e=10^{-27}~{\rm cm}^2$, beyond which caution is required, as the Earth attenuation effects would significantly suppress the BDM flux reaching underground detectors.

%%%%%%%%%%%%%%%%%%%%%%%%%%
\subsection{High Boost}
\label{subsec:highBoost}
%%%%%%%%%%%%%%%%%%%%%%%%%%
\begin{figure*}[t!]
\centering

\includegraphics[width=.495\textwidth]{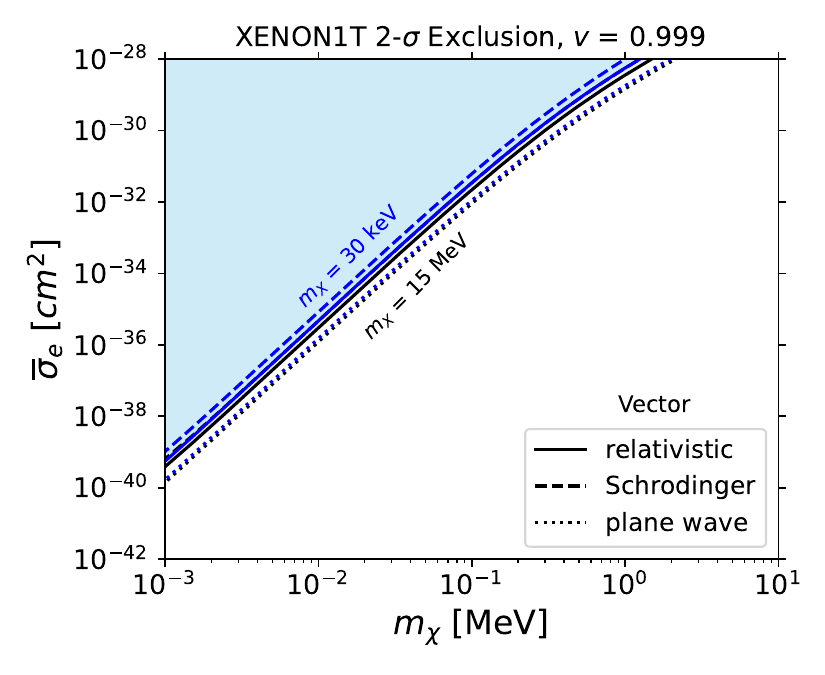}
\includegraphics[width=.495\textwidth]{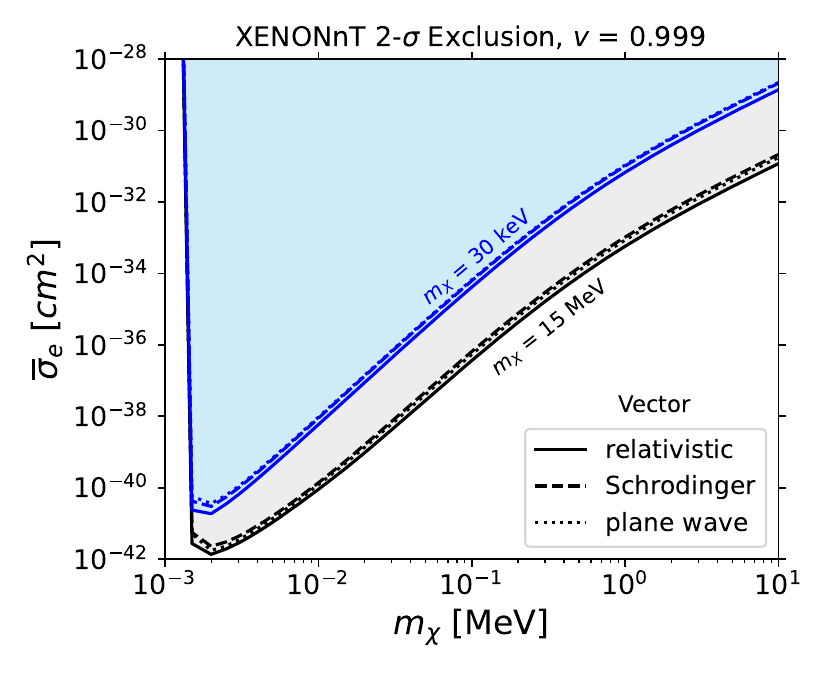}\\

\includegraphics[width=.495\textwidth]{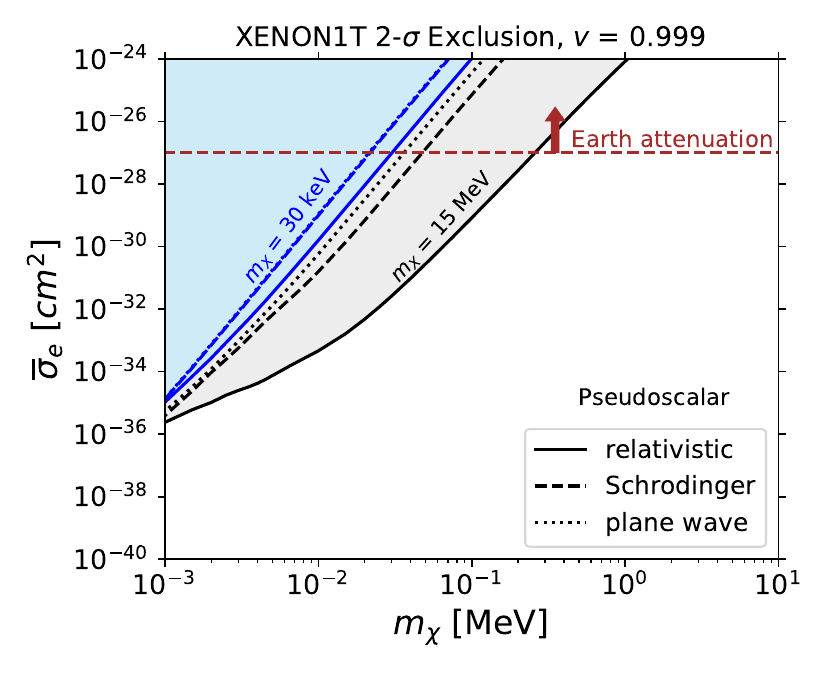}
\includegraphics[width=.495\textwidth]{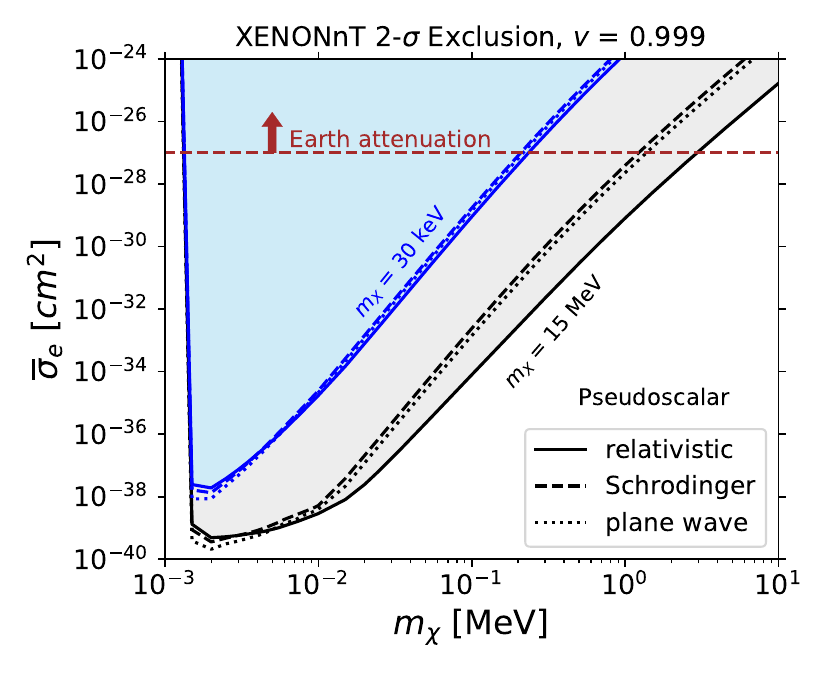}\\

\caption{$2\sigma$ exclusion limits for BDM with high boost factor ($v=0.999$) at XENON1T and XENONnT. The black lines correspond to a mediator mass of 15 MeV, while the blue lines represent a mediator mass of 30 keV. The top panel shows results for a vector mediator, and the bottom panel depicts results for a pseudoscalar mediator. The three ionization form factor treatments: DCE, SCE, and PW are represented by solid, dashed, and dotted lines, respectively. A reference cross section is indicated by the red dashed lines at $\bar{\sigma}_e=10^{-27}~{\rm cm}^2$, beyond which the Earth attenuation effects would significantly suppress the BDM flux reaching underground detectors. Therefore, interpretations in this region should be made with caution.}
\label{fig:p999_nT_1T_limits_shaded}
\end{figure*}

\begin{figure*}[t!]
\centering

\includegraphics[width=.495\textwidth]{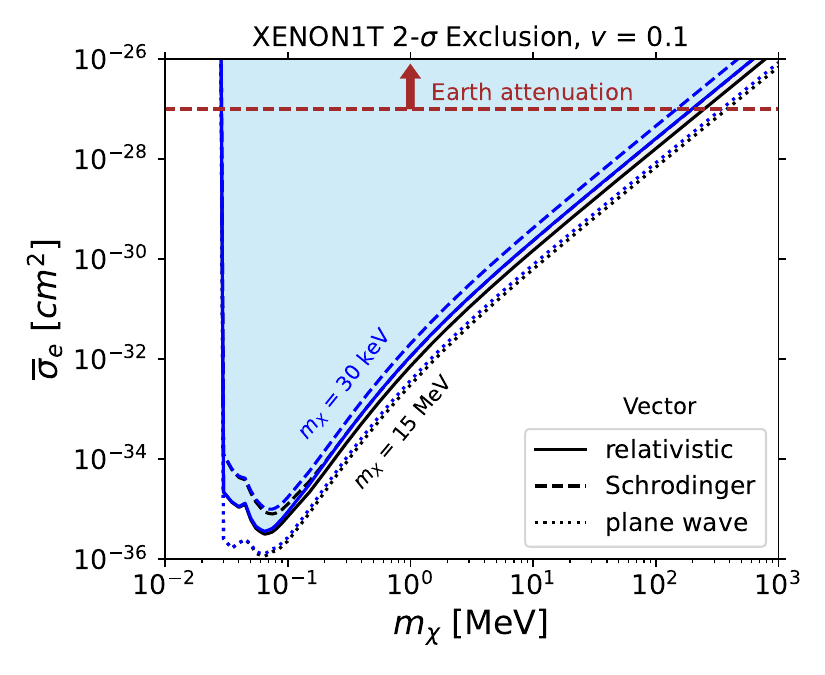}
\includegraphics[width=.495\textwidth]{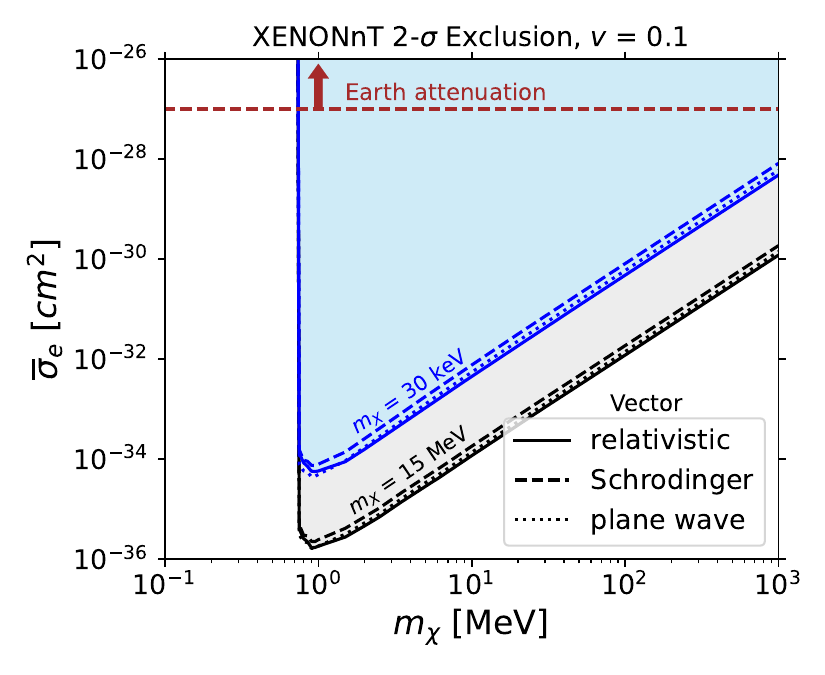}\\

\includegraphics[width=.495\textwidth]{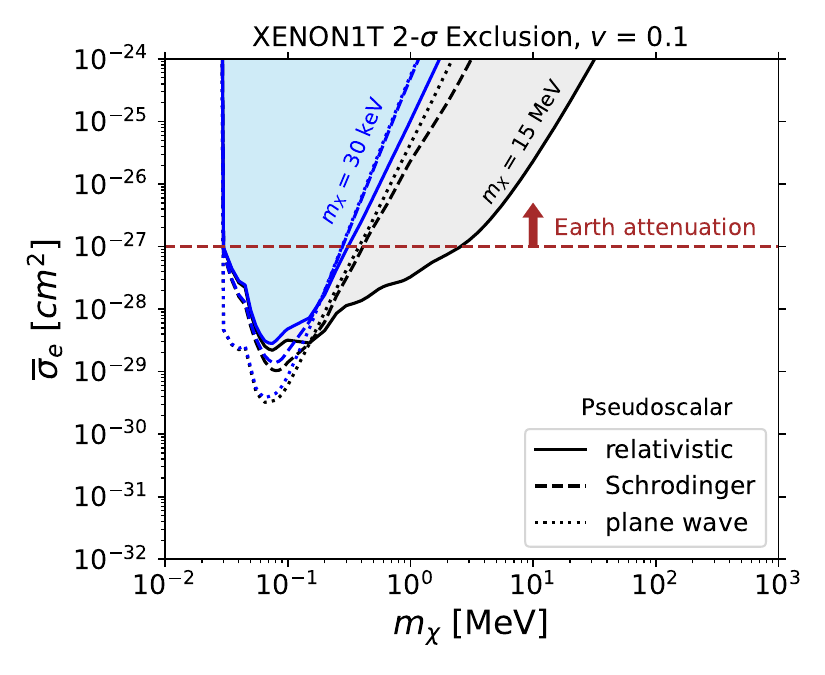}
\includegraphics[width=.495\textwidth]{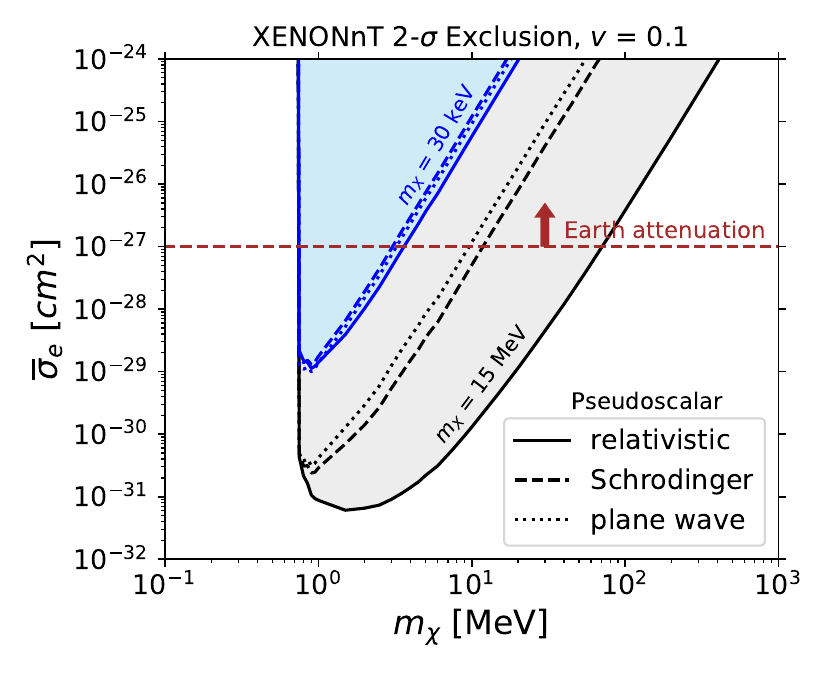}\\

\caption{$2\sigma$ exclusion limits for BDM with low boost factor ($v=0.1$) at XENON1T and XENONnT. The black lines correspond to a mediator mass of 15 MeV, while the blue lines represent a mediator mass of 30 keV. The top panel shows the results for a vector mediator, and the bottom panel depicts the results for a pseudoscalar mediator. The three ionization form factor treatments: DCE, SCE, and PW are represented by solid, dashed, and dotted lines, respectively. A reference cross section is indicated by the red dashed lines at $\bar{\sigma}_e=10^{-27}~{\rm cm}^2$, beyond which the Earth attenuation effects would significantly suppress the BDM flux reaching underground detectors. Therefore, interpretations in this region should be made with caution.}
\label{fig:p1_nT_1T_limits_shaded}
\end{figure*}

Here, we focus on the extreme boost regime, where the incoming BDM particle is highly energetic with a velocity of $v=0.999$. Figure~\ref{fig:p999_nT_1T_limits_shaded} illustrates the sensitivity estimates for the XENON1T and XENONnT experiments. 
Each panel presents $2\sigma$ exclusion curves, where the black lines with gray shading correspond to the heavy mediator, and the blue lines with light-blue shading represent the light mediator. 
The three ionization form factor treatments: DCE, SCE, and PW are depicted by solid, dashed, and dotted lines, respectively, as discussed in Section~\ref{sec:rates_ion}. 
In all cases, the exclusion limits weaken as the BDM mass increases due to a decreasing event rate.

For the vector mediator shown in the top panel, all three ionization form factor treatments yield similar limits. 
The limits for both the heavy and light mediators at XENON1T are nearly the same, making it difficult to distinguish between them, whereas XENONnT exhibits a clearer separation. 
This difference arises because the performed XENON1T analysis focuses on small electron recoils (around 1 keV and below), while the XENONnT analysis extends up to 140 keV. 
The differential cross section for the heavy mediator remains nearly flat across the electron recoil spectrum, leading to a constant event rate. 
In contrast, the cross section for the light mediator decreases with recoil energy, resulting in an exponentially falling event rate at higher energies. 
Consequently, the exclusion limits for the heavy mediator are stronger in XENONnT than in XENON1T.

A similar trend is observed for the pseudoscalar mediator in the bottom panel, except that the different ionization form factor treatments lead to significantly improved exclusion limits, sometimes differing by a few orders of magnitude. 
The non-relativistic PW and SCE treatments yield similar results, while the relativistic DCE treatment deviates substantially, highlighting its model dependence and flexibility in accommodating different Lorentz interaction structures. 
It is important to notice how significant the deviation is at XENON1T while it is mild at XENONnT. 
This must be understood in view of the boost factor of the DM particle and the capability of the two experiments. 
In this particular case, the DM particle interacts with atomically bound electrons with high speed and energy. 
Since XENONnT probes electron recoils up to 140 keV while XENON1T electron recoils are of the keV order, we observe that the atomic effects due to the ionization form factor become clearly important at low energy electron recoils.
For example, at DM mass $m_{\chi} = 0.1 $ MeV, the deviation in the limits between the relativistic and the non-relativistic treatments in XENONnT is about 1 order of magnitude while at XENON1T the deviation reaches 4 orders of magnitude.
In the next case, we will discuss this observed deviation at a much lower boost factor.

In summary, comparing the ionization form factor effects for the heavy and light mediators, we find that non-relativistic treatments provide a good approximation to the relativistic case for vector interaction. 
However, this approximation does not necessarily hold for other interaction structures, such as the pseudoscalar mediator. This model dependence arises from the structure of the radial integrals (Section~\ref{sec:rates_ion}), where wave function components remain unmixed in the vector case but are mixed in the pseudoscalar case. In what follows, we test our observations at much smaller energies by reducing the boost factor.

%%%%%%%%%%%%%%%%%%%%%%%%%%
\subsection{Low Boost}
\label{subsec:ModerateLowBoost}
%%%%%%%%%%%%%%%%%%%%%%%%%%
As the velocity of the DM decreases, its kinetic energy also decreases, making lighter DM more challenging to detect, as shown in Figure~\ref{fig:p1_nT_1T_limits_shaded}; the range of accessible mass values is reduced.
The colors, curves, and benchmark representations in this figure follow the same conventions as in Figure~\ref{fig:p999_nT_1T_limits_shaded}, except that we now consider a speed of $v=0.1$.

The overall behavior of the limit curves exhibits trends similar to those observed in the high boost case, with some differences.
First, the limit curves for the pseudoscalar mediator weaken with increasing mass at a much steeper rate compared to the vector mediator. 
This occurs because the interaction rate in the pseudoscalar case decreases more rapidly with increasing DM mass than in the vector case.
This observation can be understood semi-analytically. Assuming that the mediator mass scale is much greater than the momentum transfer scale, we find that the leading terms in the differential cross section take different forms in the vector and pseudoscalar mediator cases. In the vector mediator case, we have
\begin{equation}
\frac{d\sigma_{\rm V}}{dE_r} \propto 1+\frac{f(E_r)}{m_\chi^2},
\end{equation}
whereas in the pseudoscalar case, we find
\begin{equation}
    \frac{d\sigma_{\rm PS}}{dE_r} \propto \frac{g(E_r)}{m_\chi^2},
\end{equation}
where both $f$ and $g$ are polynomials in $E_r$. These expressions indicate that as $m_\chi$ increases, the pseudoscalar cross section continues to fall, while that of the vector case stabilizes due to the presence of the constant term. 

Second, there is a noticeable deviation in the limit curves between the relativistic and the non-relativistic treatments.
For the vector mediator, the overall deviation is minimal.
However, in the XENON1T experiment, particularly at very low masses, the deviation can reach an order of magnitude. 
This indicates that at low boost factors and small energy scales, the non-relativistic treatment is no longer a good approximation of the relativistic one, in contrast to the high boost scenario. 
This is due to the asymptotic behavior of all the ionization form factors. 

\bigskip
In general, at high boost factors and consequently high energy ($E_r$), all ionization form factors converge to a Dirac delta function, producing results similar to those expected if the electrons were initially free. However, as the boost factor and energy decrease, deviations among the ionization form factors become apparent. 
For the pseudoscalar mediator, the deviation is more pronounced.
While the model dependence plays a significant role as discussed above, the boost factor (and DM energy) also influences the results.
In the case of XENONnT, which probes higher electron recoil energies, a deviation of nearly 3 orders of magnitude is observed, whereas it was milder in the high boost case.

In summary, we find that 
\begin{itemize}
\item the relativistic and non-relativistic treatments yield different results depending on the mediator type, and 
\item even within the same mediator type, differences may arise; for example, for the pseudoscalar mediator case, the relativistic treatment enables more improved sensitivity reaches.
\end{itemize}

Finally, we note that constraints from other experiments exist for the interactions of ambient DM with electrons~\cite{SENSEI:2020dpa,EDELWEISS:2020fxc,PandaX-II:2021kai,XENON:2024znc,SuperCDMS:2020ymb,DAMIC:2019dcn,DarkSide:2018ppu,Essig:2017kqs}. Most of these constraints apply to the region $m_\chi>1$~MeV and are evaluated independently of the specific mediator type governing the DM-electron interaction. Therefore, we do not include them in our figures, as they may not be directly applicable to our scenarios.

%%%%%%%%%%%%%%%%%%%%%%%%%%
\subsection{Combined Analysis}
\label{subsec:CombinedAnalysis}
%%%%%%%%%%%%%%%%%%%%%%%%%%
\begin{figure*}[t!]
\begin{center}

\hspace{-.50cm} \includegraphics[width=.34\textwidth]{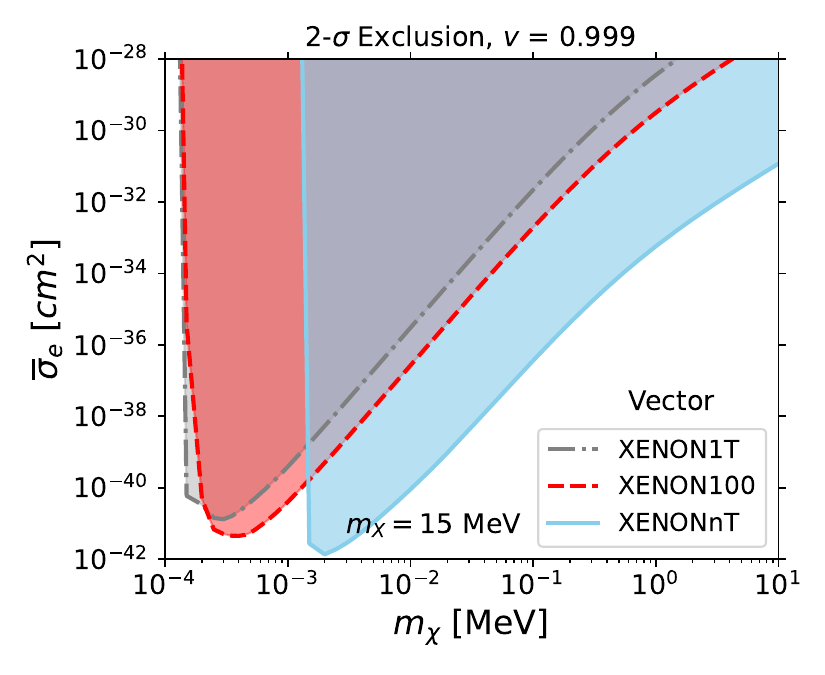}
\hspace{-.2cm} \includegraphics[width=.34\textwidth]{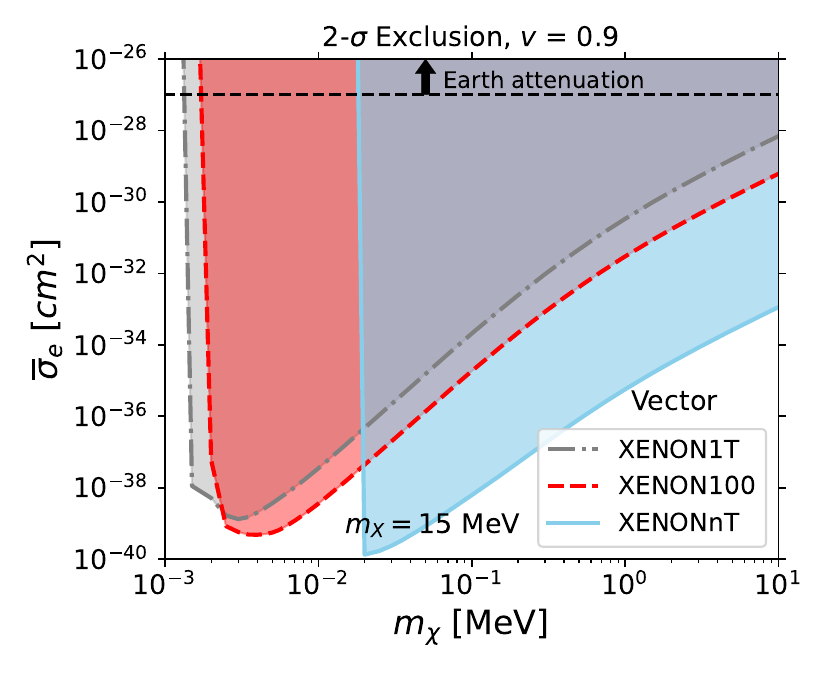}
\hspace{-.2cm} \includegraphics[width=.34\textwidth]{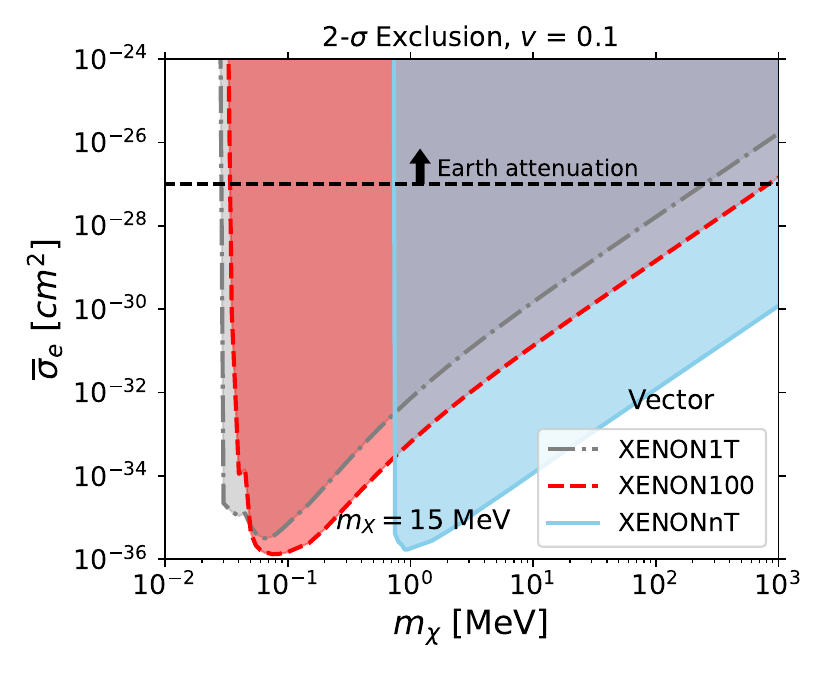}\\

\hspace{-.50cm} \includegraphics[width=.34\textwidth]{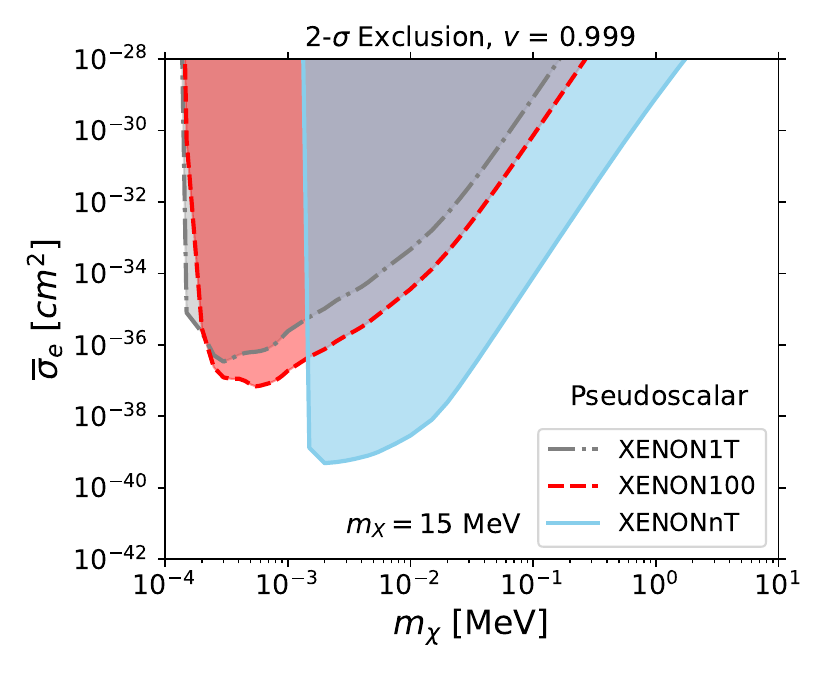}
\hspace{-.2cm} \includegraphics[width=.34\textwidth]{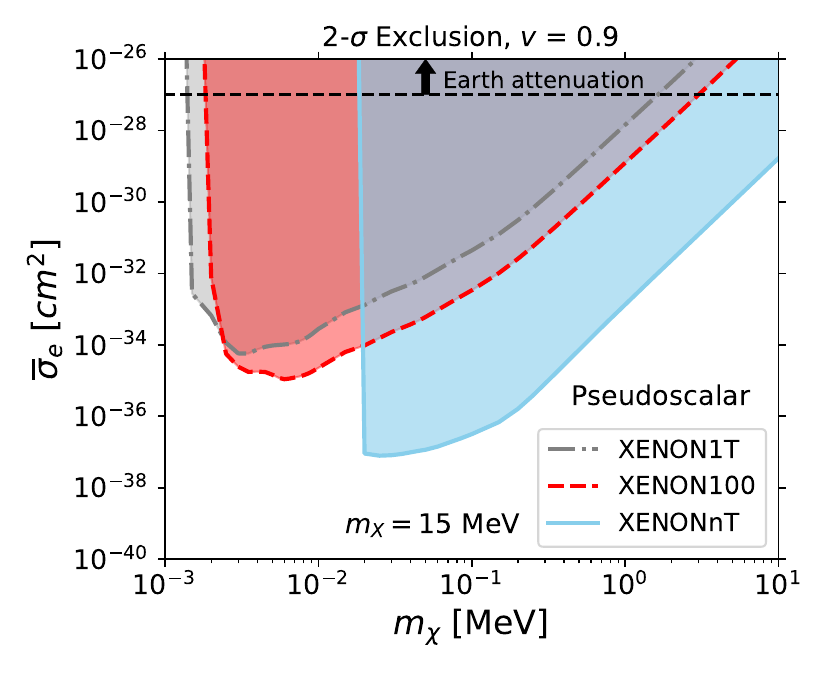}
\hspace{-.2cm} \includegraphics[width=.34\textwidth]{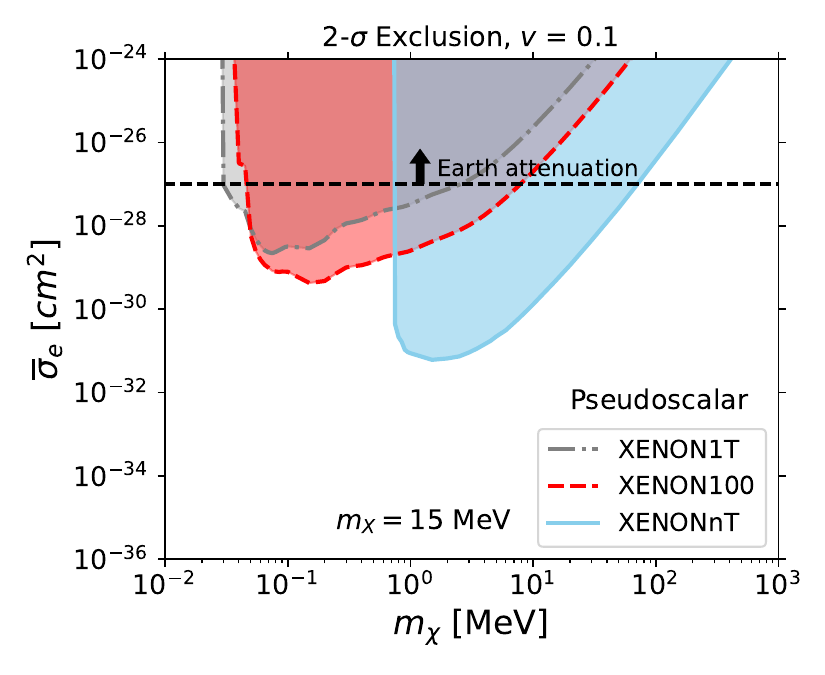}\\

\hspace{-.50cm} \includegraphics[width=.34\textwidth]{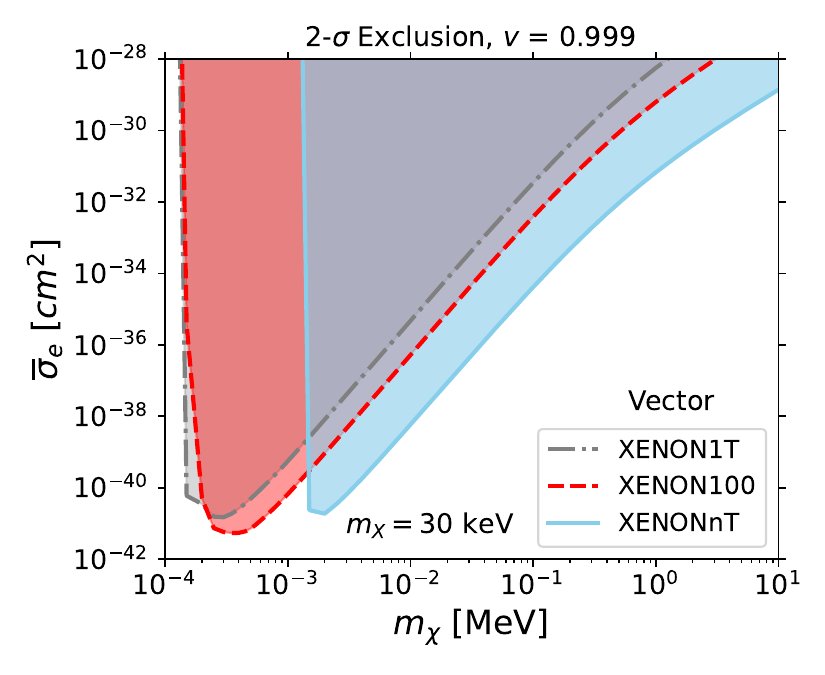}
\hspace{-.2cm} \includegraphics[width=.34\textwidth]{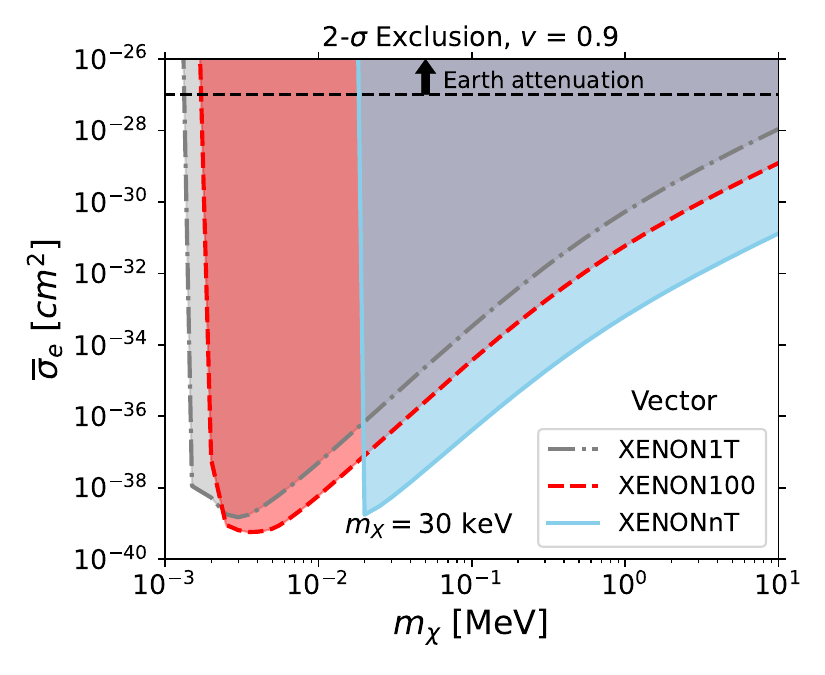}
\hspace{-.2cm} \includegraphics[width=.34\textwidth]{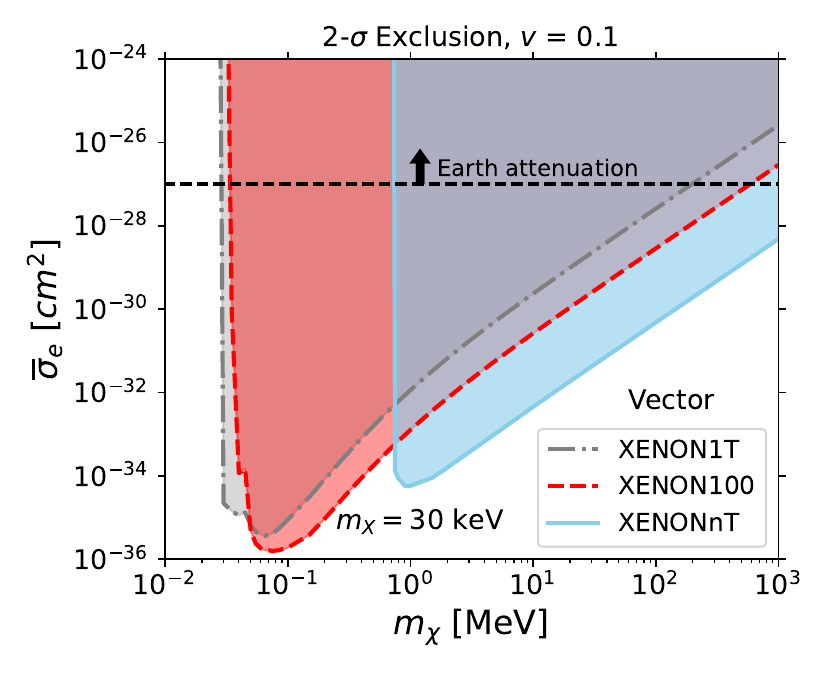}\\

\hspace{-.50cm} \includegraphics[width=.34\textwidth]{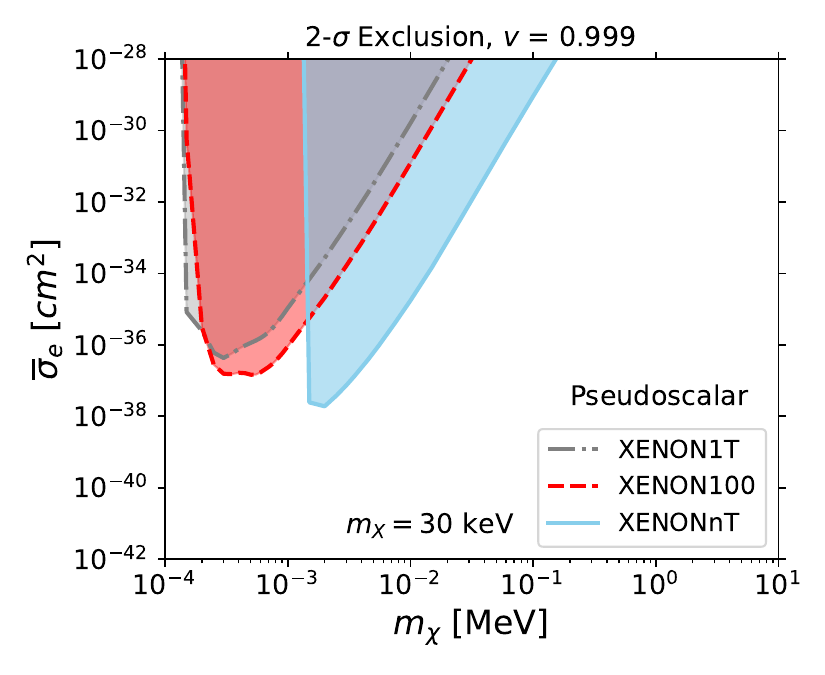}
\hspace{-.2cm} \includegraphics[width=.34\textwidth]{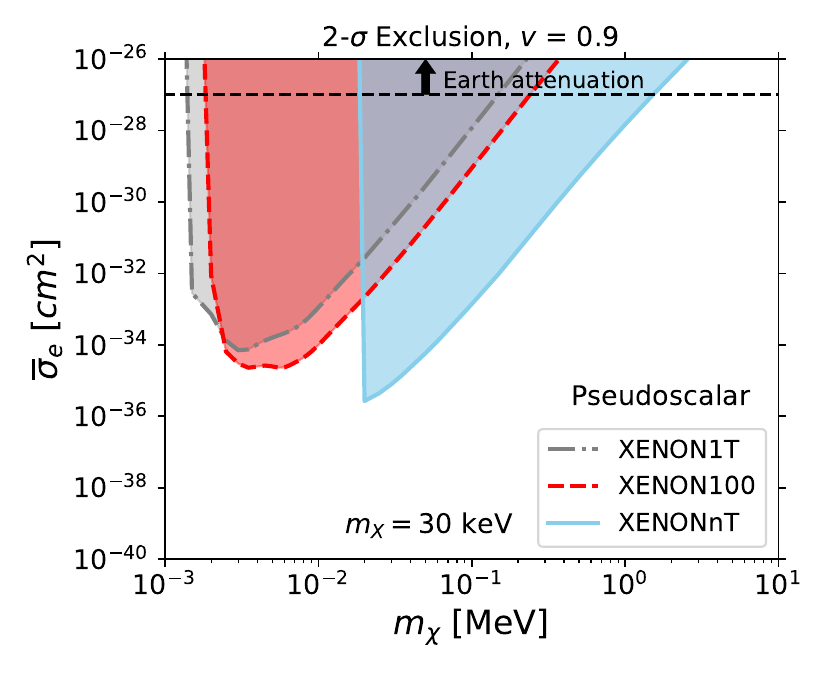}
\hspace{-.2cm} \includegraphics[width=.34\textwidth]{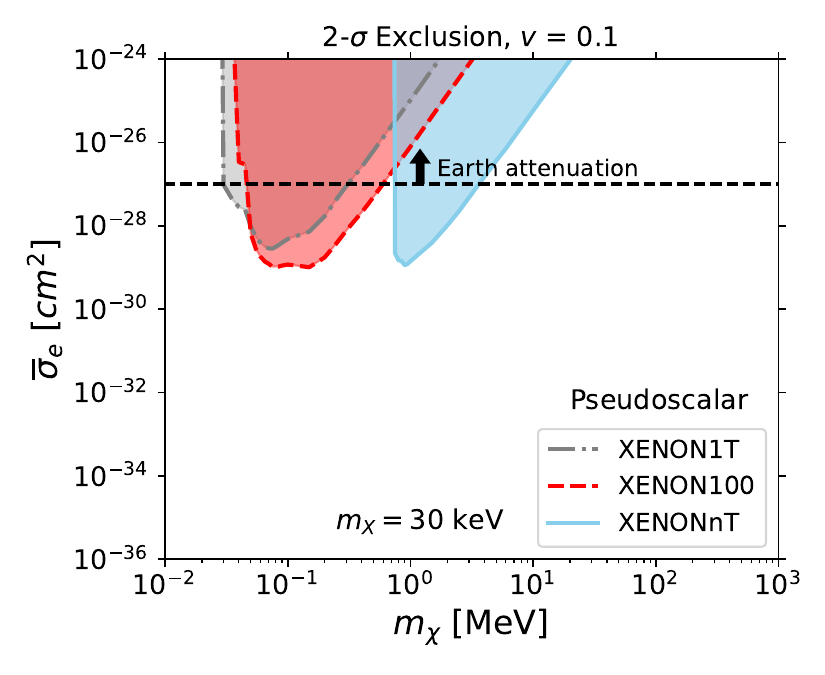}\\

\end{center}
\vspace*{-0.5cm}
\caption{2$\sigma$ exclusion limits obtained using the relativistic treatment of the ionization form factor. 
The left, center, and right columns correspond to DM speeds of $v = 0.999$, $v = 0.9$, and $v = 0.1$, respectively. 
The first two rows show the results for a heavy mediator: the vector mediator in the first row and the pseudoscalar mediator in the second. 
The last two rows correspond to a light mediator, with the vector mediator in the third row and the pseudoscalar mediator in the fourth. 
Each subplot includes three limit curves: the dot-dashed curve with a gray shaded region (XENON1T), the dashed curve with a light-red shaded region (XENON100), and the solid curve with a light-blue shaded region (XENONnT). A reference cross section is indicated by the black dashed lines at $\bar{\sigma}_e=10^{-27}~{\rm cm}^2$, beyond which the Earth attenuation effects would significantly suppress the BDM flux reaching underground detectors. Therefore, interpretations in this region should be made with caution.}
\label{fig:combined_limits}
\end{figure*}
We summarize our findings using limit plots for the three considered experiments at high, moderate, and low DM boost factors. 
These plots focus on the relativistic ionization form factor, while our earlier observations for the non-relativistic case remain valid.
Figure~\ref{fig:combined_limits} presents the 2$\sigma$ exclusion curves. 
From left to right, the columns correspond to high boost ($v=0.999$), moderate boost ($v=0.9$), and low boost ($v=0.1$). 
From top to bottom, the rows represent heavy vector, heavy pseudoscalar, light vector, and light pseudoscalar mediators. 
Each plot includes exclusion limits from XENON1T (dot-dashed gray curve), XENON100 (dotted light-red curve), and XENONnT (solid light-blue curve). 
The XENON100 limits closely follow those of XENON1T, consistent with the photoelectron (PE) spectrum analysis discussed earlier.
Overall, XENON100 and XENONnT probe smaller masses than XENON1T due to their different energy scales, with XENON1T having a slight advantage due to its lower number of detected PEs. 
The boost factor affects the probed mass scale, as indicated by the left boundary of the shaded regions. 
Higher boost factors allow access to lower DM masses, while lower boost factors do not—a purely kinematic effect. Variations in the left boundary arise from differences in experimental thresholds.
For vector mediators, the distinction between light and heavy mediators is less pronounced in XENON100 and XENON1T but becomes clearer in XENONnT. 
This results from differences in electron recoil sensitivity, with XENONnT benefiting from an extended recoil range.

%%%%%%%%%%%%%%%%%%%%%%%%%%%
\section{Conclusions}
\label{sec:conc}

In this paper, we focused on the role of atomic ionization form factors in understanding electron recoil signals induced by BDM or fast-moving DM in direct detection experiments that utilize liquid xenon targets. By comparing non-relativistic and relativistic treatments, we have highlighted the conditions under which each approach is applicable and demonstrated the importance of adopting relativistic formalism for scenarios involving significant energy transfer. 
This comprehensive approach ensures more accurate estimations of projected detection signals, emphasizing the model-dependent nature of ionization form factors in cases where DM interacts with electrons via light or heavy mediators.

Our analysis illustrates that relativistic treatments are essential for precise predictions, especially for heavy nuclei with complex atomic structures, as non-relativistic calculations can underestimate cross sections due to the behavior of Coulomb-like wave functions at small distances. Using benchmark two-component BDM scenarios with pseudoscalar and vector mediators, we have shown the mediator dependence of the results, providing insights into the interplay between the form factors and DM interaction models. We emphasize that our analysis procedures are readily applicable to other scenarios with fast-moving DM. 

This work not only refines the theoretical tools necessary for interpreting experimental data but also enhances the prospects for probing BDM and other non-conventional DM scenarios in future direct detection experiments. By identifying key factors that influence sensitivity estimates, we aim to guide the development of next-generation detectors capable of exploring uncharted parameter spaces in the search for DM. Further studies incorporating additional mediators and experimental setups will expand on these findings and strengthen the experimental reach for uncovering the nature of DM. 

%%%%%%%%%%%%%%%%%%%%%%%%%%%

\section*{Acknowledgements}
We thank Lance Dixon, Tien-Tien Yu, and Aaron Vincent for their helpful discussions.
DK acknowledges support from DOE Grant DE-SC0010813. 
KK is supported in part by the US DOE under Award No DE-SC0024407.
During the completion of this work, GM was supported in part by the UC Office of the President through the UCI CHancellor's Advanced Postdoctoral Fellowship, the U.S. National Science Foundation under Grant PHY-2210283 and the Natural Sciences and Engineering Research Council of Canada (NSERC).
The work is supported by the National Research Foundation of Korea grant funded by the Korean government (MSIT) [RS-2024-00356960 (JCP) and, RS-2020-NR055094 and RS-2025-00562917 (SS)].
SS is also supported by IBS-R018-D1.

%\appendix

\bibliography{ref}

\end{document}